\documentclass[aip,jcp,12pt,floatfix]{revtex4}

\usepackage{color}
\usepackage{braket}
\usepackage{bm}
\usepackage{amsmath}
\usepackage{amssymb}
\usepackage{amsfonts}
\usepackage{mathrsfs}
\usepackage{multirow}
\usepackage{longtable}
\usepackage{graphicx}
\usepackage{appendix}

\begin{document}


\title{The ``Simple'' Photochemistry of Thiophene} 


\author{Michael A. Parkes}
\email[]{michael.parkes@ucl.ac.uk}
\affiliation{$^1$Dept. of Chemistry, University College London, 20 Gordon St., London WC1H 0AJ, U.K.}
\author{Graham A. Worth}
\email[]{g.a.worth@ucl.ac.uk}
\affiliation{$^1$Dept. of Chemistry, University College London, 20 Gordon St., London WC1H 0AJ, U.K.}



\date{\today}

\begin{abstract}
The simple ultraviolet absorption spectrum of thiophene is investigated using a combination of a vibronic coupling model Hamiltonian with multi-configuration time-dependent Hartree quantum dynamics simulations. The model includes five states and all twenty-one vibrations, with potential surfaces calculated at the complete active space with second-order perturbation (CASPT2) level of theory. The model includes terms up to seventh-order to describe the diabatic potentials. The resulting spectrum is in excellent agreement to the experimentally measured spectrum of Holland {\em et al} [PCCP (2014) {\bf 16}: 21629]. The, until now not understood, spectral features are assigned, with a combination of strongly coupled vibrations and vibronic coupling between the states giving rise to a progression of triplets on the rising edge of the broad spectrum. The analysis of the underlying dynamics indicates that population transfer between all states takes place on a sub-100 fs timescale, with ring-opening occurring at longer times. The model thus provides a starting point for further investigations into the complicated photo-excited dynamics of this key hetero-aromatic molecule. 
\end{abstract} 


\maketitle 



\section{Introduction}
\begin{figure}[h]
    \unitlength1cm
    \begin{picture}(8,10)
    \put(-3,0){\includegraphics[scale=1.6]{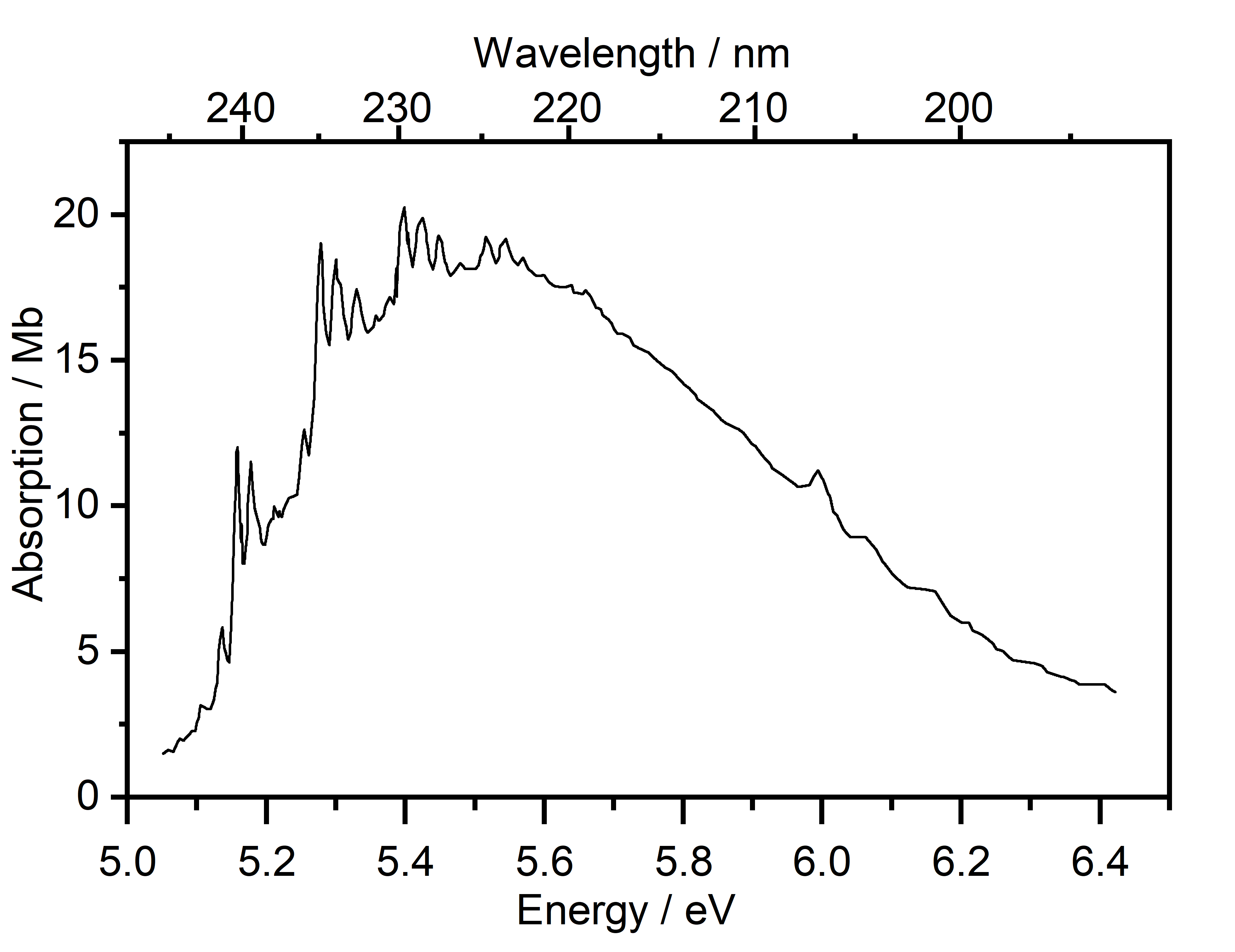}}
    \put(6,6){\framebox(2.5,2.5){\includegraphics[scale=1.4]{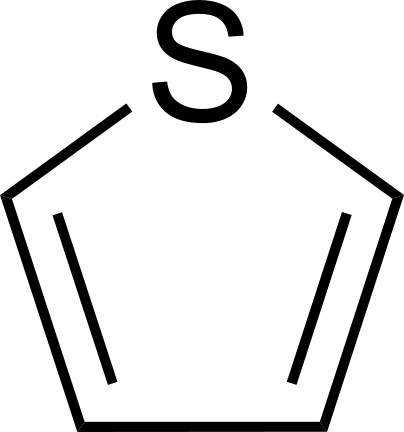}}}
    \end{picture}
    \caption{UV-vis spectrum of the first band of thiophene, taken from Holland \emph{et al} with permission of the publisher. \cite{Holland2014} The structure of thiophene is shown in the inset.}
    \label{fig:UV-Vis}
\end{figure}

Thiophene is the prototypical sulphur containing heterocycle, with a similar structure to furan but with the O atom replaced with an S atom (see inset of fig \ref{fig:UV-Vis}). Thiophene has come to play a central role in advances in optoelectronics as it is the basis of a wide range of polymers and oligomers that find use as photovoltaics, molecular switches and even light emitting diodes.\cite{Kok2020} It is the intrinsic properties of thiophene and thiophene containing polymers, such as efficient light harvesting and charge transfer, that make them such a popular choice. It is therefore surprising that thiophene's gas-phase photochemistry is actually not well understood.

A range of theoretical and experimental studies have been performed on thiophene focusing on its first excitation band, which lies in the UV and is shown in figure \ref{fig:UV-Vis}. This band is quite broad but possess a well resolved set of vibrational progressions on the rising edge. Recently, two high-resolution VUV absorption studies have been published.\cite{Holland2014,Jones2019} These works concluded that the first band is made up of two closely spaced valence states: the $^1A_1 (\pi_2-\pi_1^*)$ and the $^1B_2 (\pi_1-\pi_1^*)$. The precise ordering of these states and their energies are hard to determine and are highly dependent on type of \emph{ab initio} calculation and basis set used.\cite{Holland2014,Prlj2015,Prlj2015b} Furthermore, the vibrational progression defies clear assignment. There is some agreement that the band origin is the peak at 5.16~eV and that the vibrational structure below that is due to hot bands.\cite{Beiting1985} However, the spacing of the progressions does not match that of any known vibrational modes of thiophene, though tentative assignment has been made to $\nu_6$. In addition, the spacing within each progression is much smaller than the frequency of any of the modes of thiophene indicating that there is strong vibronic coupling between the two electronic states.\cite{Holland2014,Koppel2004}

Three experiments have looked at thiophene's excited state dynamics: one in an indirect manner, the other two directly. Wu \emph{et al} used resonance Raman spectroscopy to look indirectly at the vibronic coupling of thiophene in cyclohexane.\cite{res_raman} They observed that five symmetric A$_1$ modes and three antisymmetric (two B$_2$ and one A$_2$) modes were important when thiophene was excited into S$_1$. Their experiments suggest that dynamics in the Franck-Condon region are along the in-plane C-S and C-C ring stretches, with ring opening occurring by crossing into a B$_2$ electronic state, they also expect crossing into an A$_2$ state.

Direct experimental measurements have used UV femtosecond pump-probe spectroscopy to examine the gas-phase dynamics of thiophene.\cite{Weinkauf2008,Schalk2018} Weinkauf \emph{et al} used a 238~nm pump to excite to S$_1$ with a 276~nm probe.\cite{Weinkauf2008} By combining their experimental results with a simulation they extracted two time constants, a fast decay on the S$_1$ potential energy surface from the S$_0$ geometry (Franck-Condon point) to the S$_1$ minimum (80~fs). This decay was then followed by an ultrafast (25~fs) decay from S$_1$ to S$_3$ leading to opening of the thiophene ring. Additionally they observed a long-lived component that did not decay in the time-frame of their experiment ($>$50~ps), this was attributed to ionization of the now linear form of thiophene. Schalk \emph{et al} used a 200~nm pump to excite to S$_2$, and with a 280~nm probe they extracted similar decays to Weinkauf \emph{et al}, plus a 450~fs delay that Weinkauf did not observe.\cite{Schalk2018} They concluded that the faster decays are ring puckering motions while the slow decay might indicate a ring opening process. We note that in both experiments the experimental cross correlation was $>$160~fs making extraction of ultrafast decays difficult.

Due to these experimental difficulties in understanding the dynamics following photoexcitation several groups have attempted to use theory to gain insight. Using different theoretical methods, including TD-DFT, DFT-MRCI, CASPT2, surface hopping methods, and more \cite{Prlj2015,Cui2011,Schnappinger2017,Stenrup2012, Salzmann2008,Kolle2016, Koppel2004}. From these, some broad conclusions have been reached. Firstly, the ordering of the first two excited states: S$_1$ has $A_1$ symmetry while S$_2$ has $B_2$ symmetry. The ordering of the higher lying states has not been agreed on and again varies with theory and basis set used. From both simulations and calculation of stationary points it seems that following photoexcitation thiophene can undergo both ring-opening and ring-puckering motions. Though the relative contributions and time scales of these motions are still not agreed upon the consensus is that ring opening will be dominant. Further, there is disagreement on whether the triplet states of thiophene are involved or not. Prlj \emph{at al} performed surface hopping calculations with ADC(2) surfaces and were able to model the broad features of thiophene's first excitation band (though the vibrational progression was not present) and found no contribution from crossing to the triplet states.\cite{Prlj2015} In contrast, Schnappinger \emph{et al} used the SHARC surface hopping method with CASSCF surfaces and found that triplet states do play a role.\cite{Schnappinger2017}

Two previous studies have applied a linear vibronic coupling (LVC) model Hamiltonian to thiophene. K{\"o}ppel \emph{et al} used an LVC model and the Multiconfiguration Time-Dependent Hartree (MCTDH) method to look at the spectroscopy of furan, pyrrole and thiophene.\cite{Koppel2004} Though obtaining good agreement with experiment for furan and pyrrole, for thiophene the extensive vibrational structure was not captured correctly. In the second study the experimental work of Holland \emph{et al} was supported by use of an LVC model to determine the position of the adiabatic excitation energies of the electronic states.\cite{Holland2014} As noted in this work, a proper understanding of the absorption structure of thiophene requires additional calculations which include vibronic coupling and nuclear dynamics.

Given thiophene's importance for organo-electronics it is essential to make sure we understand the fundamental molecular properties of isolated thiophene. We have therefore applied the vibronic coupling model to construct a model Hamiltonian, but going to higher orders of expansion than used previously. MCTDH was then used to solve the time dependent Schr{\"o}dinger equation and obtain the absorption spectrum and population dynamic with an aim to understand the source of the vibrational progression and what dynamical process are occurring following photoexcitation of thiophene.

\section{Methodology}

\subsection{The Vibronic Coupling Model Hamiltonian}
\label{methods:LVC}

The vibronic coupling (VC) model uses diabatic potential energy surfaces (PES) generated from \emph{ab initio} adiabatic surfaces. 
The use of a diabatic potential removes singularities at points where two PES meet at a
conical intersection. 
These diabatic surfaces are generated by assuming that the PES are smooth and that the diabatic potential matrix ($\mathbf{W}$) can be defined as Taylor expansions around a reference point (usually the Franck-Condon region, but any convenient point can be used). 
This generates the following model Hamiltonian:
\begin{equation}
    \centering
    \mathbf{H} = \mathbf{H^{(0)}} + \mathbf{W^{(0)}} + \mathbf{W^{(1)}} + \mathbf{W^{(2)}} + ...
\end{equation}
The zeroth order matrices ($\mathbf{H^{(0)}}$ and $\mathbf{W^{(0)}}$) are diagonal and defined in the following manner:
\begin{align}
    \centering
    H_{ij}^{(0)}(Q) & =\sum_{\alpha}^{f} \left( -\frac{\omega_{\alpha}}{2}\frac{\partial^2}{\partial Q_{\alpha}^2} + \sum_{\alpha}^{f}\frac{\omega_{\alpha}}{2} Q_{\alpha}^2 \right) \delta_{ij}\\
    W_{ij}^{(0)}(Q) & = E_i \delta_{ij}
\end{align}
where $\omega_{\alpha}$ are the frequencies of vibrational mode $\alpha$,
$Q_\alpha$ mass-frequency scaled normal mode coordinates, and
$E_i$ are the energies of the electronic state at the point where the Taylor expansion is evaluated. At this point the diabatic and adiabatic states are the same by definition.

In the standard LVC model, the remaining terms in the diabatic expansion are then given by the following expressions:
\begin{align}
    \centering
    W_{ii}^{(1)}(Q) & =\sum_{\alpha}^{f}\kappa_{\alpha}^{(i)}Q_{\alpha} \\
    W_{ij}^{(1)}(Q) & =\sum_{\alpha}\lambda_{\alpha}^{(i,j)}Q_{\alpha} \\
    W_{ii}^{(2)}(Q) & =\sum_{\alpha,\beta}\frac{1}{2}\gamma_{\alpha,\beta}^{(i)}Q_{\alpha}Q_{\beta} 
\end{align}
The on-diagonal linear terms affect vibrational modes within an electronic state and the coefficients, $\kappa_\alpha^{(i)}$, represent the gradients of the \emph{ab initio} surfaces at the Franck Condon point. Due to symmetry restrictions only the $A_1$ vibrational modes of thiophene can have a non-zero $\kappa$ value. 

The off-diagonal linear coupling terms with coefficients $\lambda_\alpha^{(ij)}$ couple the different electronic states $i$ and $j$ and hence vibrations $\alpha$ with a non-zero $\lambda$ parameter are termed coupling modes. As with the $\kappa$ parameters, symmetry limits which states are coupled by which vibrations. Pairs of states will only have a non-zero $\lambda$ parameter if the product of the vibrational and state symmetries contains the totally symmetric representation. For example, in the $C_{2v}$ point group that thiophene belongs to, the states S$_1$ and S$_2$ of $A_1$ and $B_2$ symmetry, respectively, are coupled by modes with $B_2$ symmetry.

The higher-order quadratic terms ($\gamma_{\alpha \beta}^{(i)} Q_\alpha Q_\beta$) represent frequency shifts of the excited state relative to the ground state and also Duschinsky rotations where the normal modes are mixed within a state. This model describes the potential surfaces and vibronic coupling based around the Franck-Condon point and assuming that the diabatic potentials are harmonic.
In thiophene, however, the strong anharmonicity of the potential surfaces and strong coupling between the vibrational modes means that it is necessary to go to higher orders in the Taylor expansions defining the diabatic potentials, $W_{ii}$, and couplings, $W_{ij}$,
as explained in Sec. \ref{results:VCH} The symmetry rules for non-zero parameters are simple extensions of those described for the linear terms above.

\subsection{\emph{Ab Initio} Calculations}

\begin{figure*}
    \centering
    \includegraphics[width=15cm]{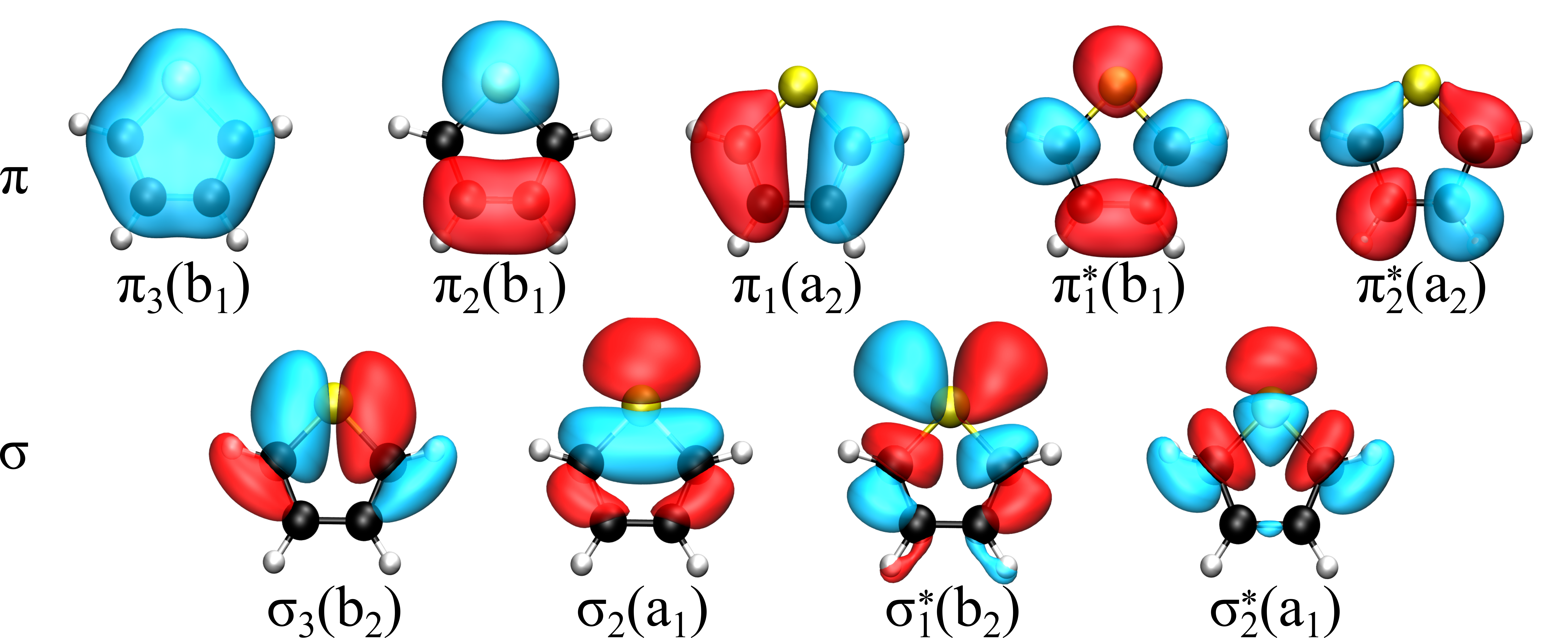}
    \caption{Molecular Orbitals included in the active space}
    \label{fig:MO}
\end{figure*}

To generate the diabatic surfaces the adiabatic singlet surfaces of thiophene were calculated at the CASPT2 level using Molpro 2015\cite{MOLPRO} with the 6-31G* basis set. The CAS space of Schnappinger \emph{et al} was used.\cite{Schnappinger2017} This consists of a 10 electron 9 orbital active space. The orbitals in the active space included all the $\pi$-orbitals and the C-S $\sigma$-orbitals. Previous work has shown that the C-S $\sigma$-orbitals must be included for correct description of the dynamics on the excited state surfaces.\cite{Schnappinger2017} See figure \ref{fig:MO} for the orbitals included in the active space. 

All calculations
used a state-averaged CAS over the lowest five singlet states (ground-state plus four excited-states), and are thus designated SA5-CAS(10,9). Test calculations were performed including a sixth excited state but this sixth state was found to be too high in energy to be important. 
The ground state structure of thiophene was optimised at the SA5-CAS(10,9) level. From the optimised structure the vibrational frequencies of all twenty one modes of thiophene were calculated, also at the SA5-CAS level. Excited-state calculations used CASPT2 (RS2C) on top of the SA5-CAS. 

The parameters for the LVC model in Sec. \ref{methods:LVC} are related to the potential gradients, Hessians and derivative coupling vectors at the Franck-Condon point, $\mathbf{Q}_0=0$
\begin{eqnarray}
\kappa^{(i)}_\alpha & = & \left. \frac{\partial V^{(i)}}{\partial Q_\alpha} \right|_{\mathbf{Q}_0}  \label{eq:kappa}\\
\gamma^{(i)}_{\alpha \beta} & = & \left. \frac{\partial^2 V^{(i)}}{\partial Q_\alpha \partial Q_\beta} \right|_{\mathbf{Q}_0} \\
\lambda^{(i,j)}_\alpha & = & F^{(i,j)}_{\alpha} (\mathbf{Q}_0)
\label{eq:lambda}
\end{eqnarray}
where $V^{i}$ is the adiabatic potential of state $i$ and 
$F^{(i,j)}_{\alpha}$ is the non-adiabatic coupling vector
between states $i$ and $j$. These quantities can thus be obtained from quantum chemistry calculations. This was done using the quantum chemistry method described above to provide the LVC model Hamiltonian.

To obtain these high-order expansion parameters, cuts through the CASPT2 potential surface were calculated at points along every normal mode and, where necessary, points along diagonal cuts between two modes. These cuts provide adiabatic surfaces for the vibronic coupling model.
Diabatic curves were then constructed that, upon diagonalisation, provide a good fit to the adiabatic energies. This fitting was performed using the VCHAM programmes \cite{vcham07} in the QUANTICS suite.\cite{Quantics2015,Worth2020} The LVC model parameters derived from Eqs. (\ref{eq:kappa}) - (\ref{eq:lambda}) were used to provide initial guess values for the $\kappa$, $\lambda$ and $\gamma$ coefficients in the fitting. 

\subsection{Quantum Dynamics Simulations}

The Multiconfiguration Time Dependent Hartree (MCTDH) method is a well established way to solve the time-dependent Schr{\"o}dinger equation for molecular systems.\cite{mey90:73} The method has been extensively reviewed 
\cite{BECK2001,worth2008,multibook2009} 
so only the main points will be covered here. For a given Hamiltonian, here the VC model of thiophene, the time-dependent Schr{\"o}dinger equation is solved by propagating the wavefunction. The MCTDH wavefunction is expanded in the following way:
\begin{equation}
    \centering
	\psi(\mathbf{Q},t) = \sum_{j_i=1}^{n_i}\ldots\sum_{j_p=1}^{n_p}
        \sum_{s=1}^{n_s} (t) \prod_{\kappa=1}^{p}\varphi_{j_\kappa}^{(\kappa)}(q_\kappa,t)
        |s \rangle
\end{equation}

Here $A_{j_i \ldots j_p}$ are the time dependent expansion coefficients and
$\varphi_{j_\kappa}^{(\kappa)}$ are time-dependent basis functions, known as single particle functions (SPFs), for coordinates $q_\kappa$. The coordinates are groups of physical
coordinates, $q_\kappa = (Q_1, Q_2, \ldots )$ to reduce the number of sets of functions
required and hence keep the expansion compact.
The SPFs are represented by time-independent primitive basis functions: in the work here all the modes of thiophene are harmonic oscillator discrete variable representations. \cite{BECK2001} The vector $| s \rangle$ represents the electronic states of the system.
The time-evolution of the expansion coefficients and SPFs are then found using variational equations of motion. The combination of MCTDH and vibronic coupling Hamiltonians has been highly successful in modeling the short-time response of a range of photo-excited molecules \cite{wor08:569}. Recent examples are 
maleimide \cite{leh20:25272} and cyclobutadiene \cite{bos21:6901}.

The computational effort, however, scales exponentially with the number of particles, $p$, as well as 
the size of the multi-dimensional grids describing the SPFs. For large
systems, the effort can be reduced by using the multi-layer MCTDH (ML-MCTDH) variant 
\cite{wan03:1289,man08:164116,ven11:044135}.
In this, the multi-mode SPFs are not directly expanded on a grid, but expanded using the
MCTDH form using lower-dimensional SPFs in  a second layer. These SPFs can be further expanded
in lower lowers, with the final layer expanded on primitive basis grids. The structure
of the ML-MCTDH wavefunction can be described by a tree graph. \cite{man08:164116} The one used for the calculations here is shown in the Appendix C, Fig. 12. The key for efficiency is constructing a tree in which strongly coupled modes are close together.

The photo-absorption spectrum of thiophene was simulated by taking the Fourier transform of the autocorrelation function of the ML-MCTDH wavefunction. 
\begin{eqnarray}
I (\omega) & \sim & 2 Re \int_0^T e^{i \omega t} C(t) \, dt \\
C(t) & = & \langle \psi^\ast (\tfrac{t}{2}) | \psi (\tfrac{t}{2}) \rangle
\end{eqnarray}
where the use of the wavefunction at $\frac{t}{2}$ is valid for real initial wavefunctions \cite{BECK2001} and allows an autocorrelation function to be obtained for twice the length of the propagation.
Corrections are applied to this transform to allow for the finite propagation time of the simulation. For this, a cosine function is applied to make sure the autocorrelation function becomes zero at the total propagation time, $T$. The correction also includes an exponential damping term ($\exp(-t/\tau)$) which allows for the finite experimental resolution. $\tau = 150$~fs was used in the following.

Calculations were converged with respect to the spectrum and diabatic state populations. The latter are obtained directly from the wavefunction coefficients related to each state. Convergence is ensured by increasing the basis set until the results do not change. This was done by selecting a maximum basis set size (shown in Fig. S3) and the increasing the number of SPFs for each node dynamically. \cite{men17:113}, making use of the natural orbital populations related to a set of SPFs by the associated reduced density matrix
\begin{equation}
	\rho^{(\kappa)}_{ij} = \sum_J{}^\prime A^\ast_{J^\kappa_i} A_{J^\kappa_j}
\end{equation}
where $A_{J^\kappa_i}$ is an expansion coefficient with the index $i$ for mode $\kappa$ and the sum is over all coefficients, except for those of the mode of interest, denoted by the prime on the summation. This is written for a single layer MCTDH form, but the form is equivalent for any node in a ML-MCTDH wavefunction. The
eigenvalues of this density matrix are the natural populations and provide a measure of importance for the SPFs in describing the wavefunction: low natural populations indicate the space spanned by the SPFs is a good basis set. In the dynamical SPF procedure, SPFs are added to keep the population of the least populated natural orbitals below a given threshold.

\subsection{Note on Labeling Electronic States}

In the following the electronic states in the adiabatic picture are labeled according to their state ordering and symmetry. I.e. $\mathrm{S_0 (A_1), S_1 (A_1), S_2 (B_2), S_3 (B_1)}$ and $\mathrm{S_4 (A_2)}$. In the diabatic 
picture states are given the letters $\tilde{X}, \tilde{A}, \tilde{B}, \tilde{C}$ and $\tilde{D}$ and these correlate with the ordered adiabatic states at the Franck-Condon point. I.e. where the molecular coordinates are 
those of the ground-state equilibrium geometry $\tilde{X} = \mathrm{S_0}, \tilde{A} = \mathrm{S_1}, \tilde{B} = \mathrm{S_2}, \tilde{C} = \mathrm{S_3}$ and $\tilde{D} = \mathrm{S_4}$.

\section{Results and Discussion}

\subsection{Vibronic Coupling Model}
\label{results:VCH}

The energies and symmetries of the five lowest states of thiophene at the CASPT2 level are given in table \ref{tbl:elec}. It should be noted that calculations at the CAS level with this active space produce a different ordering of the states. The four excited states are found to be very close in energy, spanning 0.6 eV. The S$_1$ and S$_2$ states are both bright at the Franck-Condon point.
The calculated normal modes and vibrational frequencies are listed in 
\ref{tbl:vibs} and compared to the experimental values. Images of the vibrations are given in Appendix B.

\begin{table}
    \begin{tabular}{ccccc}\hline
        State & Symmetry & CASPT2  & Transition & Experimental\\
            &           & / eV (\emph{f}) &      & Energies / eV \\
        \hline \\[-0.2cm]
        S$_0$    & A$_1$       & 0  & -      & 0              \\
        S$_1$    & A$_1$      & 5.641  &    $\pi{}_2 \rightarrow \pi{}_1^*$       &  5.64       \\
                &              &     (0.07)       &   (b$_1 \rightarrow $ b$_1$)& \\
        S$_2$   & B$_2$       & 6.070   &  $\pi{}_1 \rightarrow \pi{}_1^*$        & 5.97\\
                &              &     (0.10)       &   (a$_2 \rightarrow $ b$_1$) &  \\
        S$_3$    & B$_1$       & 6.203  &  $\pi{}_1 \rightarrow \sigma{}_1^*$      & 6.17    \\
                &              &      (0.00)      &   (a$_2 \rightarrow $ b$_2$) &   \\     
        S$_4$    & A$_2$       & 6.249  &  $\pi{}_2 \rightarrow \sigma{}_1^*$   & 6.33\\
                        &        &  (0.00)          &  (b$_1 \rightarrow $ b$_2$)  & \\\hline
    \end{tabular}
    \caption{Calculated properties of the first five electronic states of thiophene from a CASPT2 calculation with a 6-31G* basis set. The symmetries of the states, the energies (with oscillator strengths in brackets) and the main transition are given. Experimental values are vertical transition energies taken from reference\cite{Holland2014}}
      \label{tbl:elec}
\end{table}

\begin{table}
    \caption{Calculated and experimental vibrational modes of thiophene. Experimental data is taken from the NIST Chemistry Webbook\cite{nistwebn}. For vibrational character the main type of vibrational motion is indicated with the atoms involved and whether the vibration is ip = in-plane, op = out-of-plane, sym = symmetric or asym = anti-symmetric.}
      \label{tbl:vibs}
	\begin{tabular}{cc@{\hspace*{0.4cm}}ccc}\hline
	    Mode number & \multicolumn{2}{c}{Calculated}   & Experimental & Character \\[-0.05cm]
	    and symmetry &  \multicolumn{2}{c}{Frequency} & Frequency  & \\
	    & / eV &/ cm$^{-1}$ &  / cm$^{-1}$ & \\
        \hline \\[-0.2cm]
        1 b$_{1}$   &   0.055  & 444.40   & 452   &  Ring op\\
        2 a$_{2}$   &   0.071  & 572.38   & 567   &  Ring op\\
        3 a$_{1}$   &   0.074  & 602.46   & 608   &  C-S-C ip\\
        4 a$_{2}$   &   0.087  & 709.28   & 688   &  C-H op      \\
        5 b$_{2}$   &   0.090  & 730.38   & 751   &  S ip    \\
        6 b$_{1}$   &   0.091  & 736.74   & 712   &  C-H op   \\
        7 a$_{1}$   &   0.101  & 816.61   & 839   &  C-S-C ip     \\
        8 b$_{2}$   &   0.113  & 911.50   & 872   &  C ip   \\
        9 b$_{1}$   &   0.113  & 911.67   & 867   &  C-H op     \\
        10 a$_{2}$  &   0.116  & 938.83   & 903   &  C-H op     \\
        11 a$_{1}$  &   0.136  & 1100.39  & 1036  &  C-H ip     \\
        12 b$_{2}$  &   0.148  & 1193.83  & 1085  &  C-H ip     \\
        13 a$_{1}$  &   0.148  & 1194.61  & 1083  &  C-H ip    \\
        14 b$_{2}$  &   0.173  & 1398.60  & 1256  &  C-H ip      \\
        15 a$_{1}$  &   0.186  & 1504.98  & 1360  &  C ip      \\
        16 a$_{1}$  &   0.193  & 1558.35  & 1409  &  C ip    \\
        17 b$_{2}$  &   0.207  & 1676.14  & 1504  &  C ip     \\
        18 b$_{2}$  &   0.420  & 3387.27  & 3086  &  asym C-H  \\
        19 a$_{1}$  &   0.421  & 3402.02  & 3098  &  sym C-H   \\
        20 b$_{2}$  &   0.426  & 3435.98  & 3125  &  asym C-H  \\
        21 a$_{1}$  &   0.426  & 3437.96  & 3126  &  sym C-H    \\\hline
    \end{tabular}
\end{table}

The normal modes associated with the frequencies in Table \ref{tbl:vibs} form the basis for 
the vibronic coupling Hamiltonian after transforming to mass-frequency scaled normal modes.
Energies for all five states at the CASPT2 level of theory were calculated at points 
along all 21 coordinates as well as diagonal cuts between the key modes, and the VC model Hamiltonian parameters optimised to fit these energies.
The key coupling constants of the VC model are given in tables 4 and 5 in Appendix B. The full set of parameters is also provided as a Quantics operator file in the provided datasets. Only the strongest coupling parameters are listed in the tables, defined by $\kappa^{(i)}_\alpha / \omega_\alpha > 0.2$ or
$\lambda^{(ij)}_\alpha / \omega_\alpha > 0.2$. These coupling strengths relate to the shift
of a harmonic oscillator minimum from the Franck-Condon point due to the gradient at that point and thus relate to the level of excitation a mode receives on excitation.

Only the totally symmetric vibrations can have non-zero values of $\kappa$ and, as shown in Table S2, all of these modes except for the high frequency modes $\nu_{19}$ and $\nu_{21}$ have significant values. This shows that for all states the gradients of the potential surfaces at the Franck-Condon point are non-negligible and the excited-state minima for all states are significantly displaced from the ground-state equilibrium geometry. 
The coupling, $\lambda$, parameters in Table S4 show that there are a large number of significant couplings between the states. In fact, all vibrations except for the highest four frequencies are involved. 

To provide a good fit of the adiabatic surfaces for the five states of thiophene, the high-order expansion terms listed in the Appendix B Table 6 were added to the simple LVC model. These were chosen according to the shapes of the potentials and the symmetry of the modes involved. I.e. the product of the symmetries of the modes and states involved in a term has to be totally symmetric. The diabatic potentials along most modes for most states include a fourth order parameter to account for symmetric anharmonicity away from the Franck-Condon point, with modes 3 and 7 requiring a sixth order term to keep them bound. The modes with $A_1$ symmetry can also take a third-order parameter to increase the asymmetry of the potentials. 

The final model has a total of 707 parameters. For the final Hamiltonian, 2967 points were included in the fit. The overall root mean square deviation (RMSD) of the energies at the points from the quantum chemistry and the model is 0.39~eV. This drops to 0.21~eV if only points with energies below 8~eV are considered. This is not an
insignificant error, but the restricted form of the potentials means that some regions, mostly away from the Franck-Condon point are not well fitted while by 
inspection the key regions around the Franck-Condon point look to be well described. 

\begin{figure}
    \centering
    \includegraphics[width=12cm]{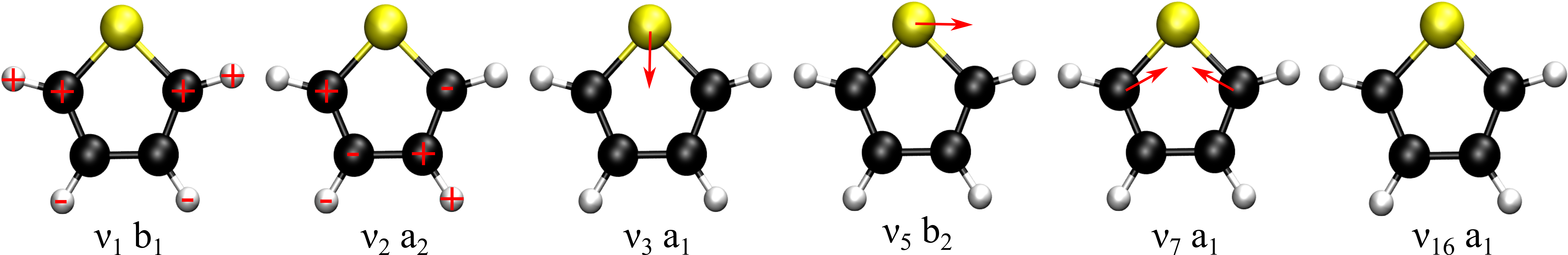}
    \caption{Key normal modes $\nu_1$, $\nu_2$, $\nu_3$, $\nu_5$ and $\nu_7$}
    \label{fig:vibs}
\end{figure}

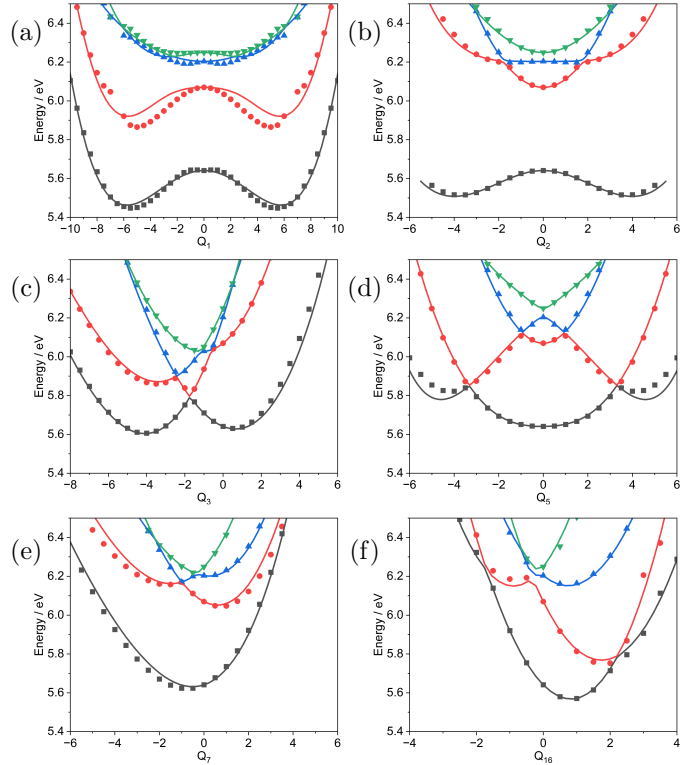
\begin{figure}
\unitlength1cm
    \begin{picture}(8,10)
    \put(0,0){\includegraphics[width=9cm]{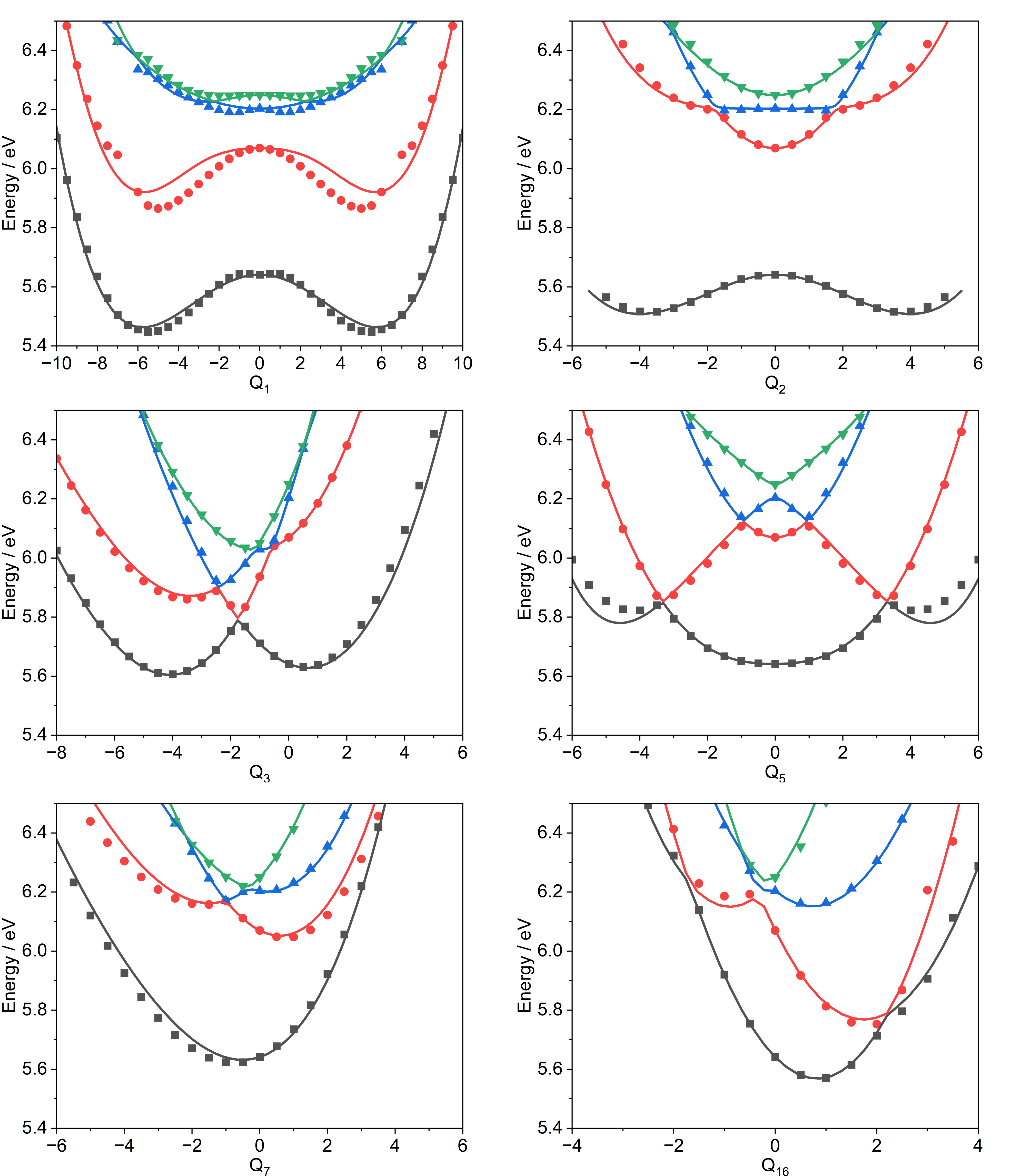}}
    \put(-0.1,9.7){\makebox(0,0)[c]{\footnotesize (a)}}
    \put(4.4,9.7){\makebox(0,0)[c]{\footnotesize (b)}}
    \put(-0.1,6.3){\makebox(0,0)[c]{\footnotesize (c)}}
    \put(4.4,6.3){\makebox(0,0)[c]{\footnotesize (d)}}
    \put(-0.1,2.8){\makebox(0,0)[c]{\footnotesize (e)}}
    \put(4.4,2.8){\makebox(0,0)[c]{\footnotesize (f)}}
    \end{picture}
    \caption{Cuts along selected mass-weighted normal modes for thiophene. (a) Q$_1$ (b) Q$_2$ (c) Q$_3$ (d) Q$_5$ (e) Q$_7$ and (f) Q$_{16}$.  Only states S$_1$ to S$_4$ are shown}
    \label{fig:key_cuts}
\end{figure}

From the optimised parameters, in most states high-order terms provide coupling between modes $\nu_1$, $\nu_2$ and $\nu_3$. Modes $\nu_3$, $\nu_5$, and $\nu_7$ are also strongly coupled. 
The key vibrations are thus $\nu_1, \nu_2, \nu_3, \nu_5$ and $\nu_7$ and these vibrations are show in figure \ref{fig:vibs}.
    The cuts though the fitted potential energy surfaces along these key modes are shown in figure \ref{fig:key_cuts} with cuts along the pairs of modes in \ref{fig:diag_cuts}. Cuts along the remaining modes are shown in Appendix B.3. 

The totally symmetric vibrations $\nu_3$ and $\nu_7$ are responsible for the symmetric sulphur motion and ring compression, respectively. The cuts along these modes in figure \ref{fig:key_cuts}(c) and (e) show the close nature of the four excited states. The cut along $\nu_3$ is particularly dramatic, with the $\tilde{C}$ and $\tilde{D}$ diabatic
    potentials falling sharply to end up below the $\tilde{A}$ and $\tilde{B}$ states at negative displacements, with a double well on the adiabatic S$_1$ state resulting. While it is not found to be important for the description of the spectrum, and it does not couple strongly to the other modes, the cut along the totally symmetric mode $\nu_{16}$ is also shown in Fig. \ref{fig:key_cuts} (f). It is a ring stretch and there is a low lying crossing between S$_1$ and S$_2$ (diabatic states $\tilde{A}$ and $\tilde{B}$) to positive displacement.
    
    The b$_2$ mode $\nu_5$ represents the asymmetric sulphur motion across the ring. The cuts along this mode in Fig \ref{fig:key_cuts}(d) also show a large deviation from harmonic behaviour of states $\tilde{C}$ and $\tilde{D}$, with the $\tilde{C}$ state again dropping rapidly in energy and ending up below the $\tilde{A}$ state at large displacements and forming minima on the adiabatic S$_1$. Modes $\nu_1$ and $\nu_2$ with $b_1$ and $a_2$ symmetry, respectively, are out-of-plane carbon motions. They both form double well potentials
    for all the excited states (Figs \ref{fig:key_cuts}(a), (b)), but an energy gap is retained between S$_1$ and the higher excited states.

    The cuts between modes that describes the coupling between vibrations show strong coupling. The cut Q$_{3-5}$ between $\nu_3$ and $\nu_5$ (Fig. \ref{fig:diag_cuts} (d)) is particularly dramatic, with the S$_1$ minimum to large negative displacement dropping over 0.5 eV below the curves seen along the cuts in Fig. \ref{fig:key_cuts}(c) and (d).
    The mixing of $\nu_3$ with the non-symmetric $\nu_1$ and $\nu_2$ vibrations (Figs \ref{fig:diag_cuts} (b), (c)) provides a complicated structure to the potential crossings that is hard to fit. The potential curves along the $\nu_1 / \nu_3$ diagonal (Fig. \ref{fig:diag_cuts} (b)) are particularly poor, unable to fully capture the strong anharmonicity to large negative coordinates for states S$_2$ and S$_3$.

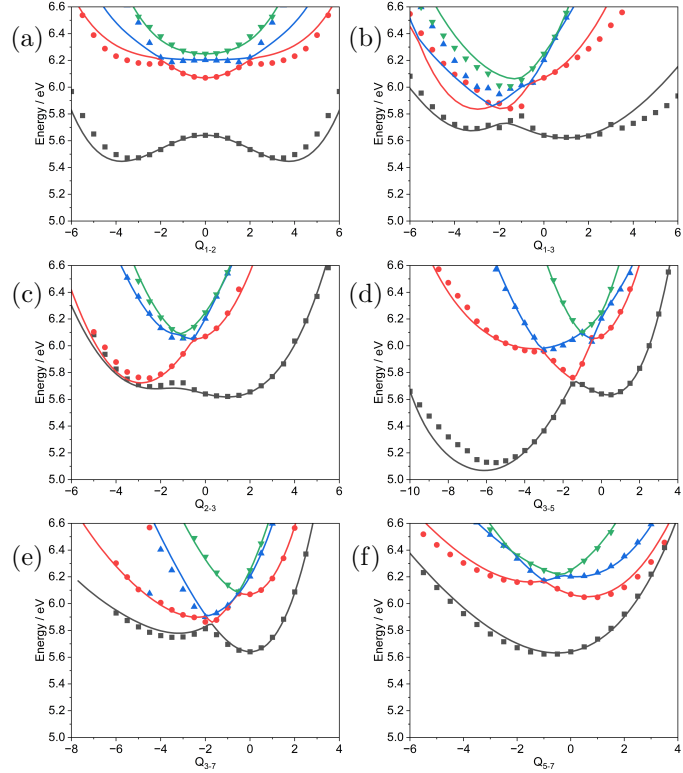
\begin{figure}
\unitlength1cm
    \begin{picture}(8,10)
    \put(0,0){\includegraphics[width=9cm]{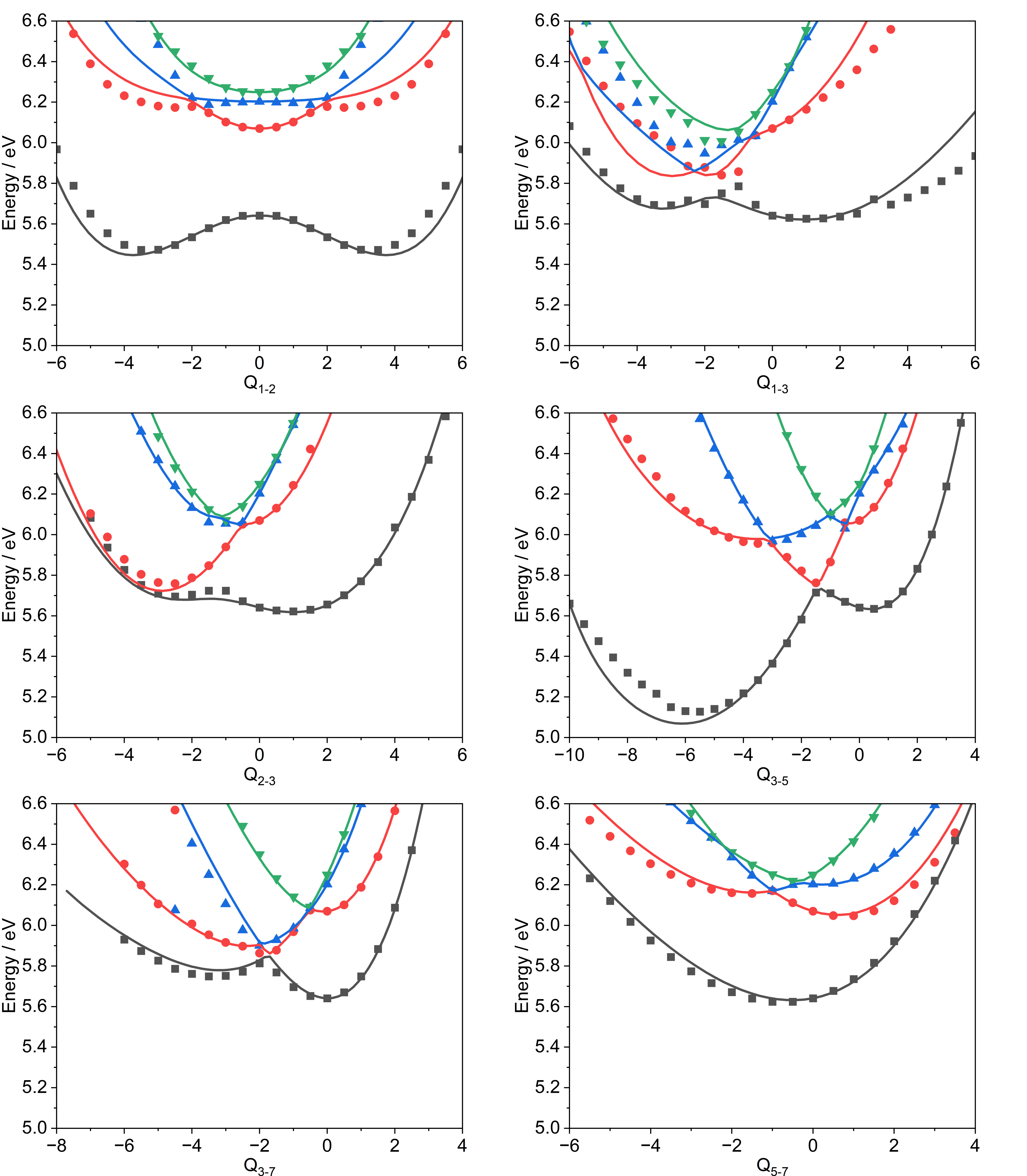}}
    \put(-0.1,9.7){\makebox(0,0)[c]{\footnotesize (a)}}
    \put(4.4,9.7){\makebox(0,0)[c]{\footnotesize (b)}}
    \put(-0.1,6.3){\makebox(0,0)[c]{\footnotesize (c)}}
    \put(4.4,6.3){\makebox(0,0)[c]{\footnotesize (d)}}
    \put(-0.1,2.8){\makebox(0,0)[c]{\footnotesize (e)}}
    \put(4.4,2.8){\makebox(0,0)[c]{\footnotesize (f)}}
    \end{picture}
    \caption{Cuts along selected mass-weighted combinations of normal modes for thiophene. The pair of modes are indicated in the x-axis label. Only states S$_1$ to S$_4$ are shown}
    \label{fig:diag_cuts}
\end{figure}

Compared to  the earlier vibronic coupling model of K\"oppel et al \cite{Koppel2004}, the new model not only goes to higher orders in the diabatic surfaces, but the ADC(2) electronic structure calculations used in the earlier work placed the A$_2$ state just 0.1 eV above the lowest excited A$_1$ singlet state, leading to a rather different model. 

\subsection{Absorption Spectrum}

\begin{figure}
    \centering
    \includegraphics{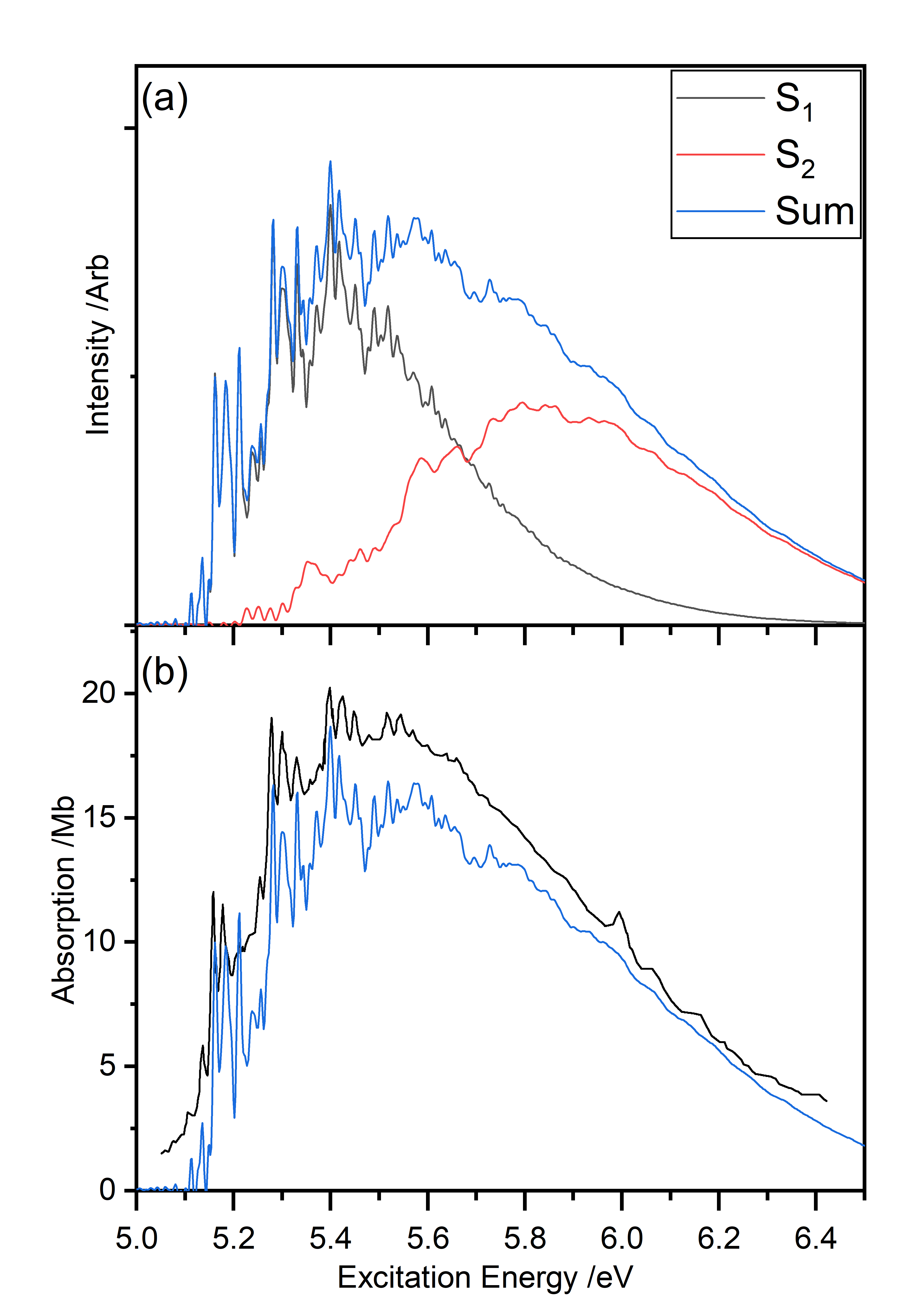}
    \caption{(a) calculated photo-absorption spectra for excitation from the S$_0$ to the S$_1$ and S$_2$ states of thiophene with all modes included in the model. (b) the experimental photo-absorption spectrum of thiophene taken from Holland \emph{et al} with the sum of the two calculated spectra. The calculated energies have been shifted by -1.85~eV.\cite{Holland2014}}
    \label{fig:spec_comp}
\end{figure}

Time dependent wavepacket calculations were performed using the ML-MCTDH method on the five state twenty-one mode vibronic coupling model. Propagations were run for 200~fs starting with a vertical excitation of the ground-state vibrational wavefunction to either the bright S$_1$ or S$_2$ states. This provides an
autocorrelation function out to 400~fs.
The calculated photo-absorption spectra obtained from the Fourier Transform of the autocorrelation functions are shown in figure \ref{fig:spec_comp}(a), both separately and summed. In (b) the summed spectrum is compared the experimental spectrum of Holland \emph{et al}.\cite{Holland2014} The calculated spectra have been shifted by 1.85~eV. As the zero-point energy of the model is 1.90~eV, the calculated vertical excitation energies are in excellent agreement with the experimental ones.

The model Hamiltonian clearly reproduces the features on the rising edge of the photo-absorption spectrum very well. The spacing of the progression on the rising edge agrees excellently between calculation and experiment and the width is also in good agreement.
The exponential damping function of 150~fs used to reduce edge artefacts when calculating the is equivalent to convoluting spectral lines by a Lorentzian of 9~meV, i.e. negligible in this broad spectrum. The broad featureless underlying spectrum is thus due to the vibronic coupling between states as this governs the lifetimes and hence spectral broadening. The model thus correctly accounts for this coupling. The simulation even reproduces the features at the onset of the spectrum which have previously been assigned as hot bands, which means they are not.\cite{Beiting1985}
The only features missing are the structure at around 6~eV. This feature is probably due to a Rydberg state of thiophene that was not included in the \emph{ab initio} calculations.\cite{Holland2014}

To confirm the quality of the fit, a 7~Hz high pass Fourier transform filter was applied to both the calculated absorption spectrum and the experimental spectrum of \citeauthor{Holland2014}. \cite{Holland2014} Applying this filter removes the broad featureless spectral component and leaves just the structure. The results of this are shown in the supplemental information and the agreement is excellent.

Given the similarity of the calculated and measured absorption spectra, the VC Hamiltonian clearly provides a good model of the early-time dynamics of thiophene after photo-excitation, capturing the state energies around the Franck-Condon point and the frequencies of the vibrations and the couplings between them. In the next sections, the model will be used to help assign the spectral features and understand how the couplings between the modes and states lead to the observed spectrum. 

\subsection{Assigning the spectrum}

\begin{figure}
    \unitlength1cm
    \begin{picture}(8,18)
    \put(-1,12){\includegraphics{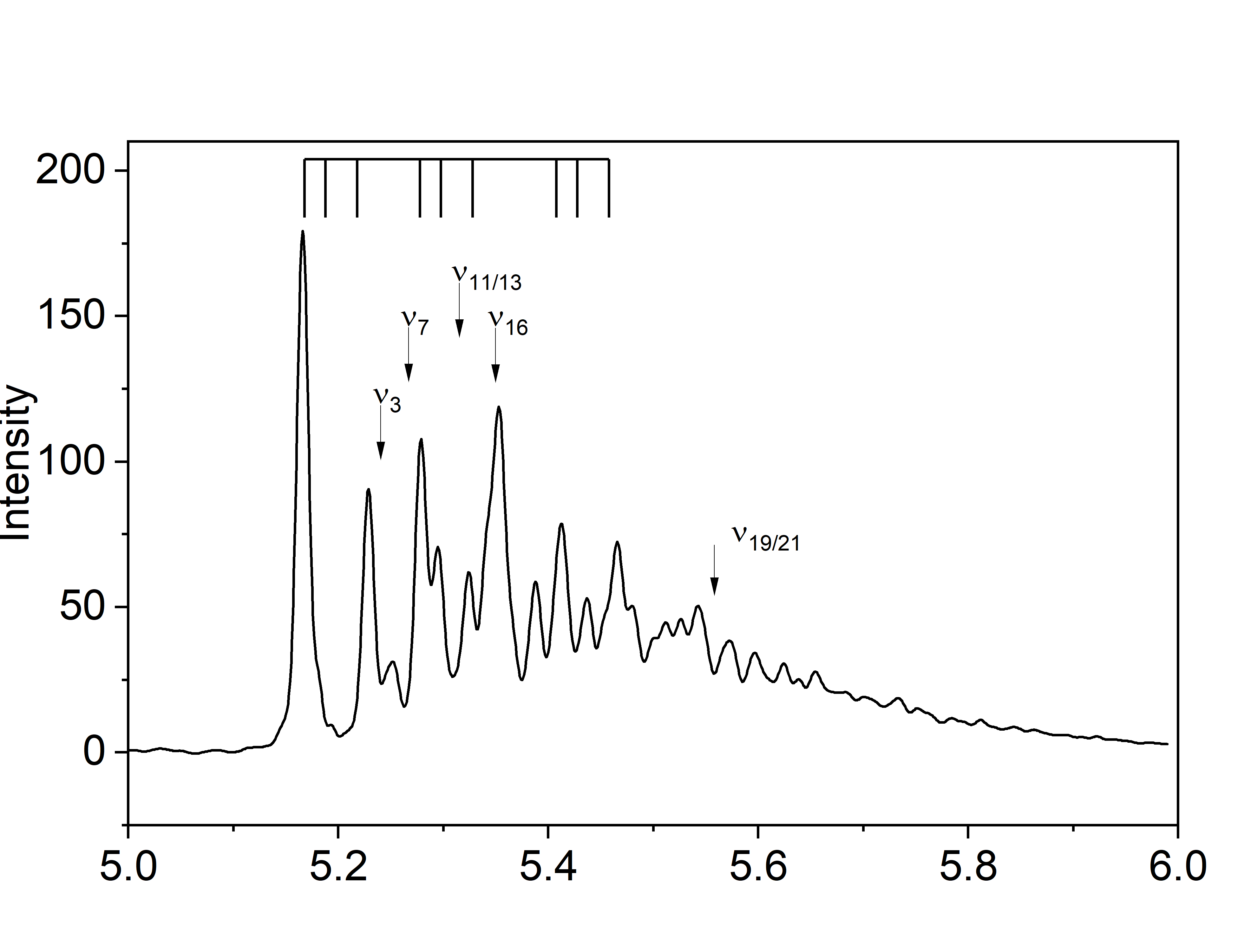}}
    \put(-1,6){\includegraphics{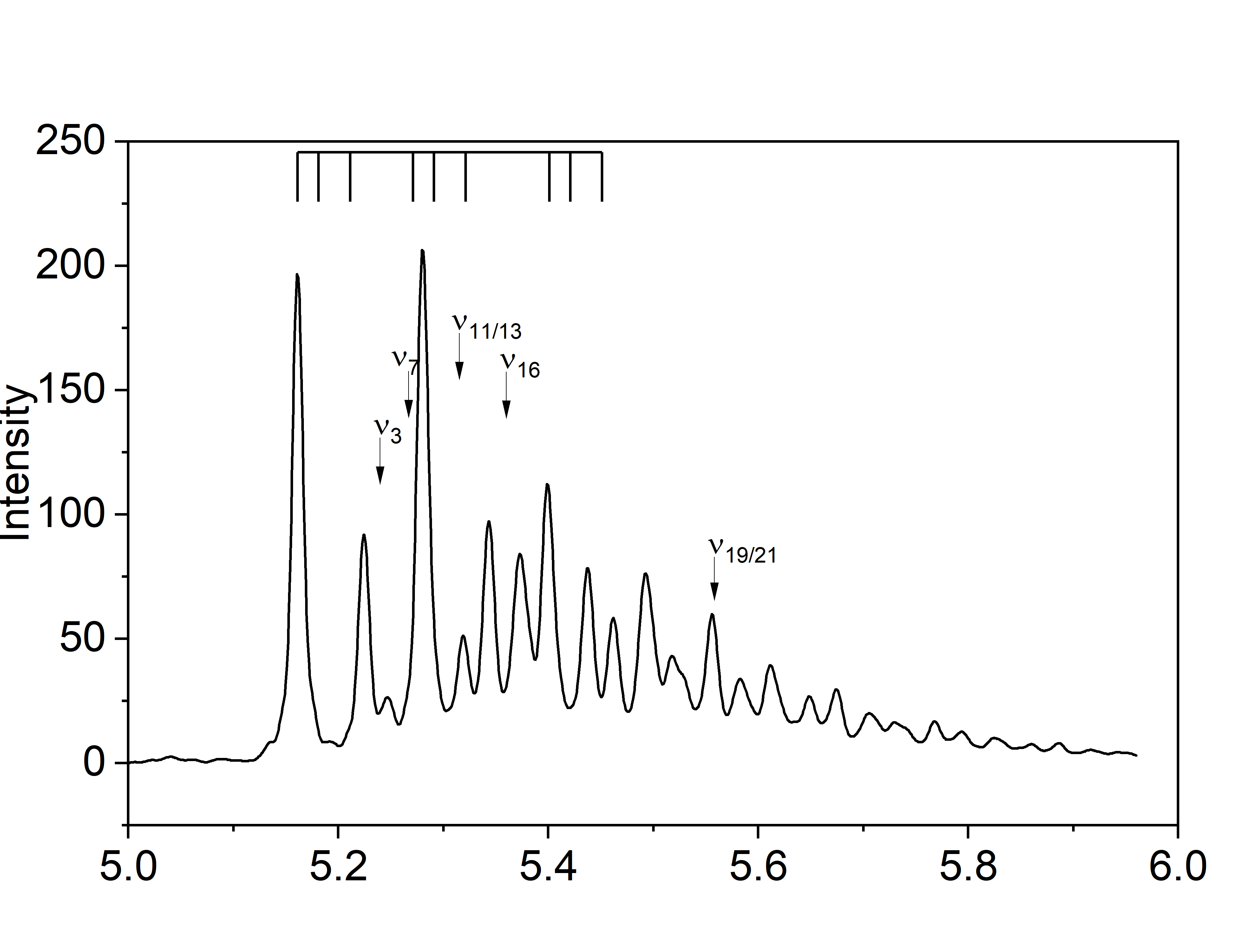}}
    \put(-1,0){\includegraphics{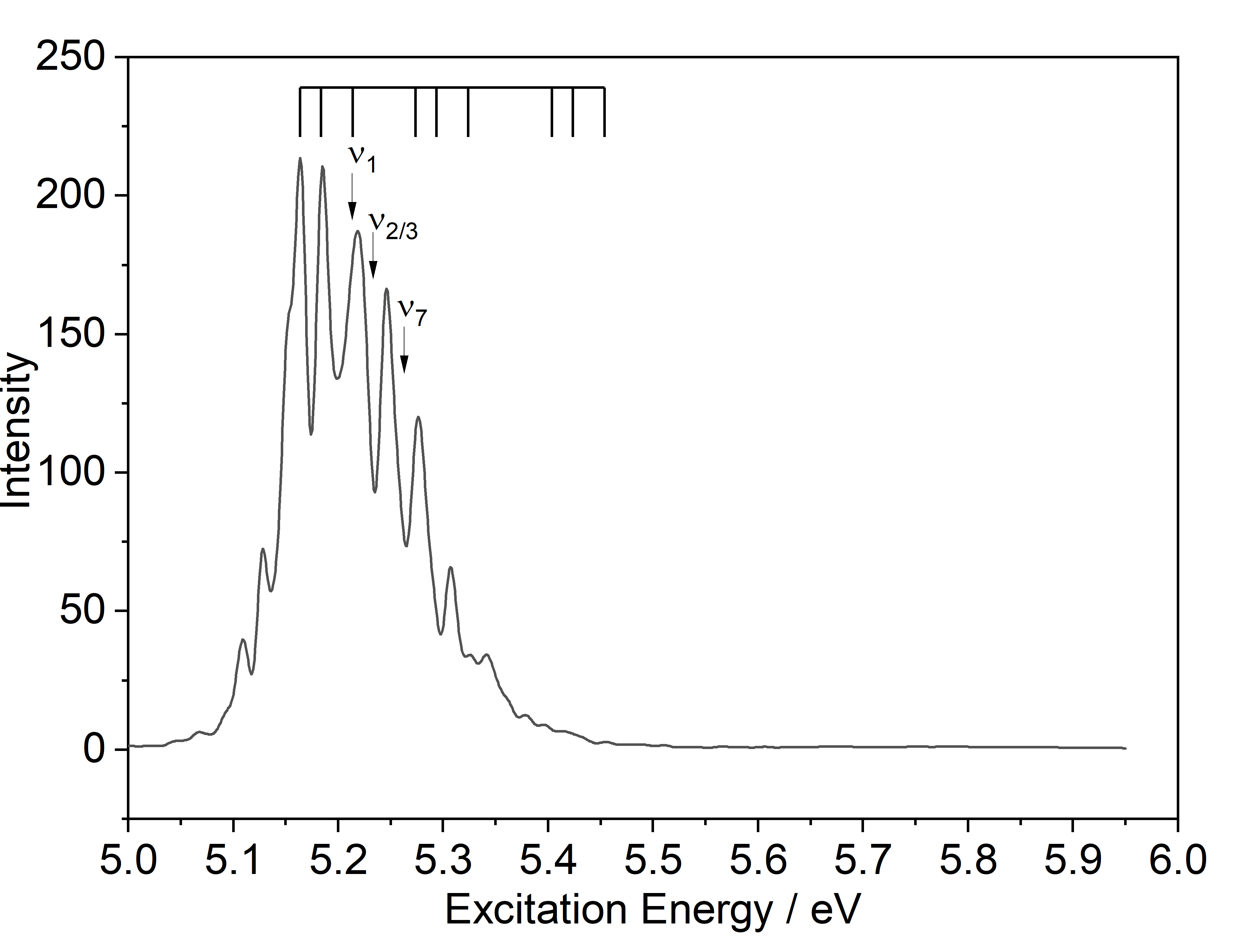}}
    \put(0,17){\makebox(0,0)[l]{\footnotesize (a)}}
    \put(0,11){\makebox(0,0)[l]{\footnotesize (b)}}
	\put(0,5.5){\makebox(0,0)[l]{\footnotesize (c)}}
    \end{picture}
	\caption{Calculated photo-absorption spectrum of thiophene for excitation into the $\tilde{A}$ state including only: (a) All totally symmetric modes, and only on-diagonal linear ($\kappa$) terms. 
	(b) All totally symmetric modes but inter-state coupling terms between $\tilde{X}$ and $\tilde{A}$ were also included along with the on-diagonal linear terms. (c) Only modes $\nu_1$, $\nu_2$, $\nu_3$ and $\nu_7$ with all relevant parameters.
    The combs marks where the experimental vibrational progressions are found. The arrows indicate the positions of the vibrational fundamentals. The spectra have been shifted by -1.01~eV -1.04~eV and -0.55~eV respectively to line up with the onset of the experimental progression} 
    \label{fig:FCF}
\end{figure}
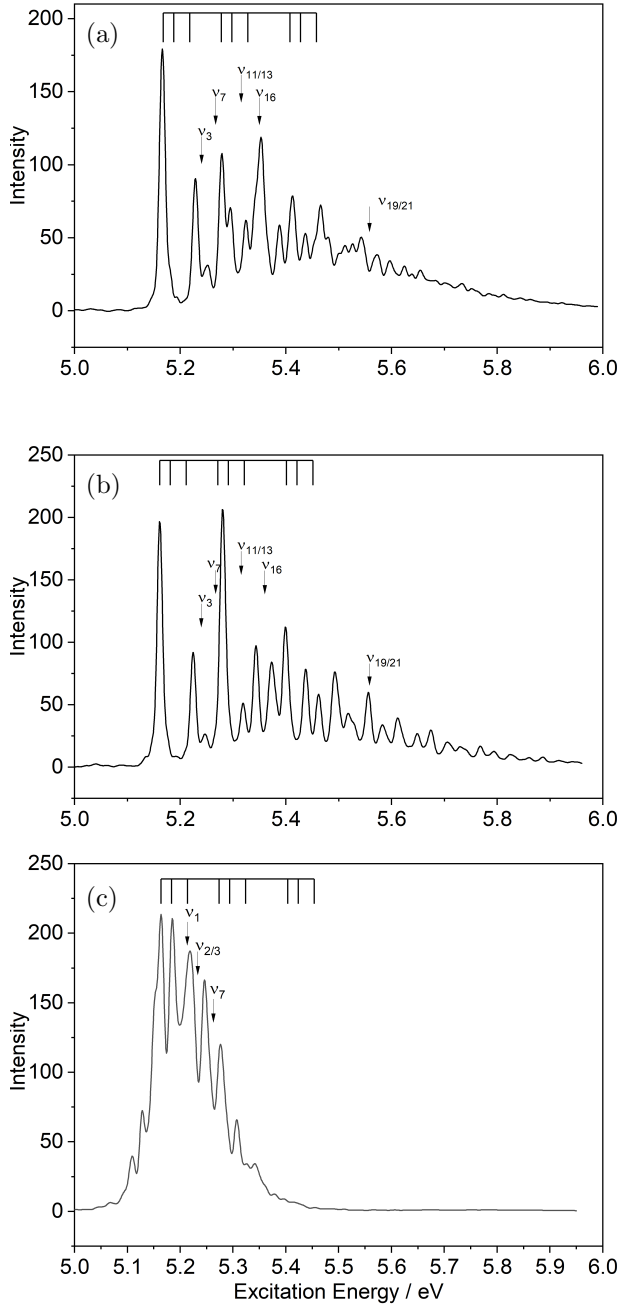

The spectrum can be assigned by running calculations including subsets of modes and couplings to see what gives rise to the features observed.
Absorption spectra are usually dominated by frequencies from the totally symmetric vibrations, which in the absence of vibronic coupling are the only modes that can be excited after absorbing a photon.
In figure \ref{fig:FCF}(a) the calculated photo-absorption spectrum of thiophene is shown for excitation from the ground state into $\tilde{A}$ using a Hamiltonian including only the eight a$_1$ modes and the on-diagonal $\kappa$ (\emph{intra}-state) coupling terms. This is the spectrum that would be produced by a calculation of the Franck-Condon factors for the transition. The comb shown above the spectrum represents the frequencies of the progressions on the rising edge of the spectrum. There are triplets, three peaks separated by 0.02~eV and 0.03~eV, which are part of progression separated by a frequency of 0.12~eV. It is clear that this spectrum does not match the experimental spectrum: the peaks are near the fundamental frequencies (shown by arrows), which bear no relationship to the progressions observed. 
Vibrational and vibronic coupling is thus the key to modeling the photo-absorption spectrum of thiophene.

Fig. \ref{fig:FCF}(b) shows the changes to the Franck-Condon spectrum if the inter-state ($\lambda$) coupling terms between $\tilde{X}$ and $\tilde{A}$ are included along with the on-diagonal linear terms for the a$_1$ mode calculation. As both states have the same symmetry (A$_1$) these couplings are only non-zero for the a$_1$ vibrations, and have particularly large values for $\nu_3, \nu_7, \nu_{11}, \nu_{13}$ and $\nu_{16}$. Again, this vibronic coupling model does not capture the full behaviour of thiophene following photoexcitation. However, the vibronic coupling has had a significant effect on the spectrum. 
The $\nu_{16}$ peak, which is the most intense in the Franck-Condon spectrum, has reduced in intensity and split.
More importantly, modes $\nu_{11}$ and $\nu_{13}$ are now degenerate and give rise to a single intense peak at the frequency of the progression of the triplets. 
Previous work \cite{Holland2014} tentatively suggested that the progression is due to excitation of the $\nu_{11}$ mode ($\nu_6$ in that paper) on the grounds of symmetry. We confirm this assignment and can explain the large
drop in frequency of approximately 0.015 eV (120 cm$^{-1})$. We also assign this to be a compound peak mixing in the $\nu_{13}$ mode.

Finally, a model Hamiltonian was used including the four strongly coupled low frequency vibrational modes. These are $\nu_1$, $\nu_2$, $\nu_3$ and $\nu_7$ and the model included all couplings involving these vibrations. Modes $\nu_1$ and $\nu_2$ are out of plane ring vibrations, while modes $\nu_3 $ and $\nu_7$ are in-plane bends of the C-S-C bond. These normal modes are shown in Fig. \ref{fig:vibs}. In Fig. \ref{fig:FCF}(c) the photo-absorption spectrum calculated using this four mode 5-state Hamiltonian is shown. The calculation captures a key feature seen in the experimental results, namely the vibrational triplet seen throughout the spectrum. The spacings of this progression matches the spacings of the previously unassigned progression on the rising edge of the experimental spectra. 
This calculation allows us to assign the structure as due to both inter-state and intra-state coupling between these four modes. It is difficult to disentangle the role of the many couplings involved and it is possible only to say that the strong vibrational and vibronic coupling involving these modes significantly changes the potential surfaces from a harmonic, or even near harmonic, form.

\subsection{Diabatic Population Dynamics}

\begin{figure}
    \centering
    \includegraphics[width=7.5cm]{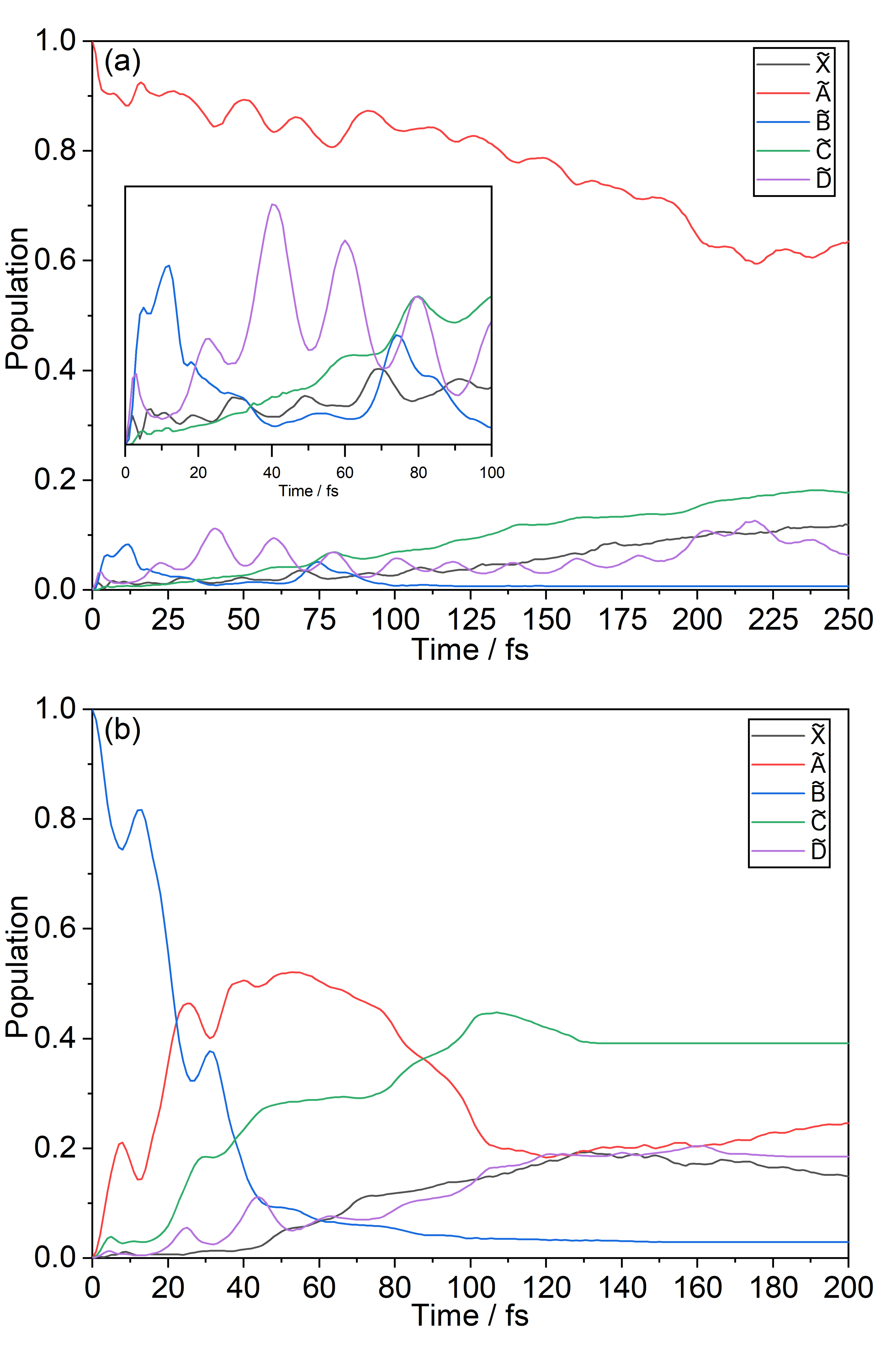}
    \caption{Diabatic state populations of thiophene following excitation into (a) the S$_1$ state using the full 21 mode Hamiltonian. The insert shows the early time populations of the initially unpopulated electronic states 
    (b) the S$_2$ state using the full 21 mode Hamiltonian.}
    \label{fig:pops}
\end{figure}

From the wavepacket propagation it is possible to extract the dynamics of the wavepacket motion on the diabatic thiophene surfaces. 
The change in diabatic state populations with time for initial excitation into the $\tilde{A}$ state (equivalent to S$_1$ at the Franck-Condon point) are shown in fig \ref{fig:pops}(a). To check whether the population is 
reaching a plateau the propagation was extended to 250 fs.
The population in $\tilde{A}$ shows a constant decay rate and does indeed reach a plateau at this time, with a population of  around 0.6. Population is transferred to all four of the other states. As shown in the inset, most of the population is initially transferred into $\tilde{B}$ and $\tilde{D}$. This population in $\tilde{B}$ has largely decayed away after 20~fs, with a brief recovery in population around 70~fs. Between 20 - 70~fs most of the transferred population is in $\tilde{D}$ which, after peaking around 40~fs, decays away to remain at a near constant value from about 80~fs until a small recovery around 200~fs. In contrast the populations in $\tilde{X}$ and $\tilde{C}$ show a slow constant increase in population, both reaching about 0.1 by 250~fs. 

$\tilde{D}$ clearly plays a key role in how the population moves between electronic states from $\tilde{A}$. The initial transfer into $\tilde{B}$ has a period of around 21~fs. This is equivalent to a frequency of 1570~cm$^{-1}$ (0.19 eV). This is presumably due to the $\nu_{16}$ vibration, which leads to a low lying crossing between $\tilde{A}$ and $\tilde{B}$ (Fig. \ref{fig:key_cuts} (f)). This period is also seen as strong oscillations in the $\tilde{D}$ state populations, indicating that the population is decaying from $\tilde{B}$ by crossing to $\tilde{D}$.
Changes in population in $\tilde{A}$ are also mirrored in $\tilde{D}$: a peak in the oscillating signal on $\tilde{A}$ is matched by a trough in $\tilde{D}$ population. Further, between 100 - 180~fs the drop in $\tilde{D}$ population is matched by a revival in $\tilde{A}$ population, with a revival in $\tilde{D}$ from 200-250 fs matched by 
another drop in $\tilde{A}$. This indicates there is also direct transfer between $\tilde{A}$ and $\tilde{D}$. 

It is interesting to compare these theoretical results to the experimental analysis of Wu \emph{et al}.\cite{res_raman} From their resonance Raman spectroscopy they expected an initial transfer of population from S$_1$ to a B$_2$ and an A$_2$ electronic state, from our calculations $\tilde{B}$ is a B$_2$ electronic state while $\tilde{D}$ is an A$_2$ state. Therefore, the population transfer we observe agrees with these Raman results. There has been a previous ultrafast pump-probe spectroscopy measurement where S$_1$ is initially populated by Weinkauf \emph{et al}.\cite{Weinkauf2008} They observed an 80~fs decay, attributed to structural relaxation on the S$_1$ excited state surface, followed by a fast decay (25~fs) leading to ring opening, there was also a long-lived component (\textgreater{}50~ps). The experimental time-scales are not directly comparable with those seen in the simulation as the experiments excite a single eigenstate rather than creating a diabatic wavepacket. However, the structural relaxation could correspond to the transfer from $\tilde{A}$ to $\tilde{B}$, which stays on the S$_1$ state and results in ring-puckering. The ring-opening, as will be examined in more detail below, is then due to crossing to the $\tilde{C}$ and $\tilde{D}$ states.

The change in diabatic state populations with time after initial excitation into the $\tilde{B}$ state (equivalent to S$_2$ at the Franck-Condon point) are shown in \ref{fig:pops}(b). The dynamics are very different to that in the $\tilde{A}$ state. In this case, the population in $\tilde{B}$ rapidly decays, with a time constant of around 20~fs, ending up with a population of just 0.05 in $\tilde{B}$ after 100~fs. The population is transferred from $\tilde{B}$ initially into $\tilde{A}$, and at later times into $\tilde{C}$ and $\tilde{D}$ and $\tilde{X}$. At around 100~fs the $\tilde{A}$ population crosses into $\tilde{C}$ and at longer times most of the population (ca 40\%) is in $\tilde{C}$ with the remaining population divided equally between $\tilde{X}$, $\tilde{A}$ and $\tilde{D}$. As with excitation into $\tilde{A}$ there is a visible oscillation in the populations of the states, this time with a spacing of about 16~fs, a frequency of approximately 2100~cm$^{-1}$ (0.26 eV). As there is no obvious vibration of this frequency able to mediate the population transfer, and this is seen directly after excitation, this must be a Rabi-like oscillation with the lower state and relates to an effective energy gap between the states. 

The time-resolved experiments by Schalk {\em et al} \cite{Schalk2018} used a pump wavelength that would excite into the $\tilde{B}$ state. They divided the observed dynamics into two regions. The first had two decay times of 25~fs and 400~fs. The second had a rise of 25 fs followed by an 80~fs decay similar to that seen in the lower energy pump experiments of Weinkauf. This 80~fs decay time they also attribute to ring-puckering and, while they did not see the 25~fs subsequent fast decay, they saw a slow 450~fs decay which they attributed potentially to ring-opening. The first region can be clearly related to the decay of the $\tilde{B}$ state in Fig. \ref{fig:pops}(b), which has a fast initial decay followed by a slow component. The rise at the start of the second region is the population flow into the $\tilde{A}$ state and the 80~fs decay is then again decay from the $\tilde{A}$ state.  

\subsection{Structural Dynamics}

The cuts through the potential surfaces in Figs. \ref{fig:key_cuts} and \ref{fig:diag_cuts} show the one-dimensional energy minima for the various diabatic states. The energy ordering and relative positions of the minima indicate that there are a number of possible minima on both the S$_1$ and S$_2$ adiabatic surfaces, with the one-dimensional minima part of a minimum energy channel. 

To find the actual lowest energy points on the surfaces, energy minimisation calculations were run starting from geometries with combinations of the one-dimensional minima coordinates. 
For the S$_1$ state, along $Q_1$ the minimum has a value of 5, while along $Q_2$ the value is 2. 
Along $Q_3$, there are two minima at 1 and -4, while along $Q_5$ there are three minima at 0 and $\pm 5$. Q$_7$ stays around 0. Four calculations were run ($Q_5 = \pm5$ are symmetrically equivalent) and two distinct minima found.
The coordinates of the minima are given in the Appendix B.4.

\begin{figure}
    \unitlength1cm
    \begin{picture}(8,8.2)
    \put(-6,-9.2){\includegraphics[scale=0.8]{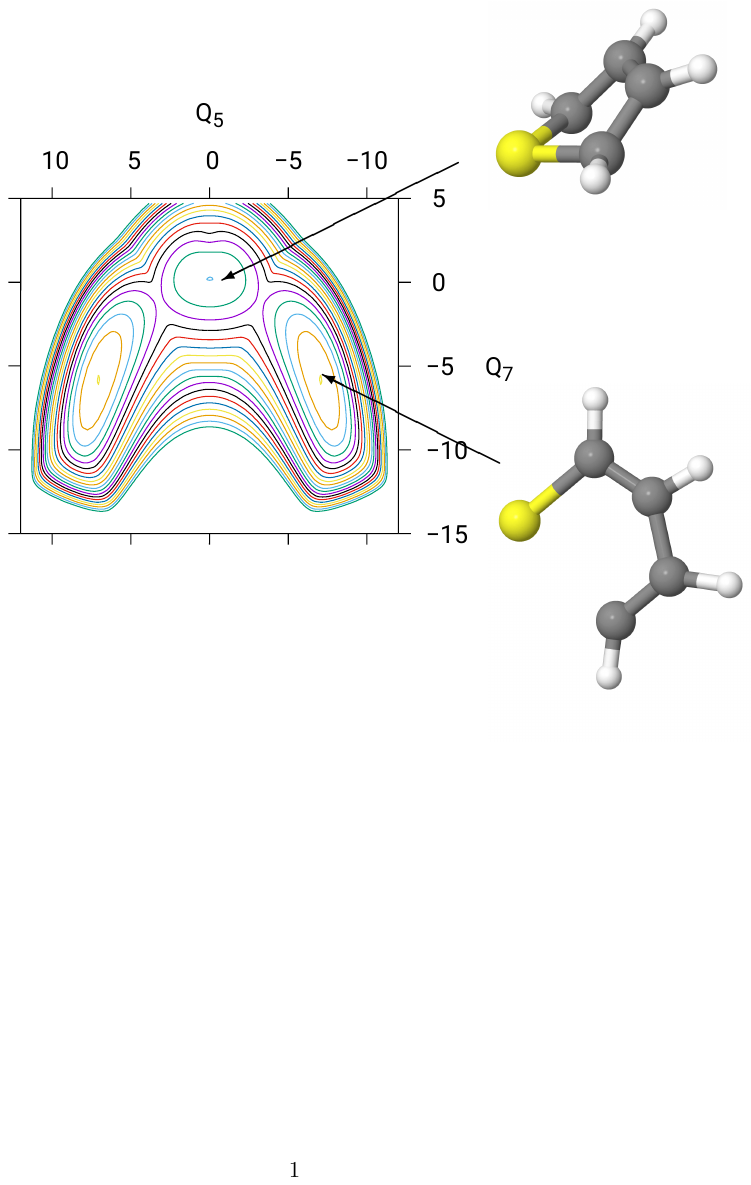}}
    \end{picture}
    \caption{Cut through the S$_1$ adiabatic potential surface along modes $\nu_5$ and $\nu_7$.
	All other coordinates have a value of zero, except Q$_1$ which had a value of 5.0.
    The structures associated with each minimum are also plotted.}
    \label{fig:s1_minima}
\end{figure}

One minimum is at 5.23 eV. It is a ring-puckered structure close to the Franck-Condon point, but with the sulphur out-of-plane. The second minimum is much lower in energy at 2.77 eV. 
This is a ring-opened structure, with a large displacement along Q$_7$. It should be noted that the structure is planar.
These minima are shown in a two-dimensional cut through the adiabatic S$_1$ potential in the space of Q$_5$ and Q$_7$ in Fig. \ref{fig:s1_minima}. The two channels leading to the minima are clear, and the minimum energy structure for each channel is shown alongside. 

The structures fit with the correlation of the minima to the diabatic states. The ring puckered structure with small displacements from the Franck-Condon point comes from the $\tilde{A}$ diabatic state, which is the $\pi \pi^\ast$ excitation, whereas the ring-open structure with large displacements along Q$_3$, Q$_5$ and Q$_7$ correlates to the $\tilde{C}$ diabatic state with its $\pi \sigma^\ast$ character.

A similar search on the S$_2$ adiabatic surface found four distinct minima. Three, at 5.34, 5.50 and 5.68~eV, are ring puckered structures close to the Franck-Condon point, while the fourth is again a low energy (3.95 eV) structure which is an in-plane ring-open structure similar to that in Fig. \ref{fig:s1_minima}. It is not straightforward to associate the ring-puckered structures with particular diabatic states, but the ring open structure clearly correlates to the $\tilde{D} (\pi \sigma^\ast)$ diabatic state.

As a result of this analysis, transfer of diabatic state population from
states $\tilde{A}$ and $\tilde{B}$ to $\tilde{C}$ and $\tilde{D}$ can be considered as moving from ring-closed to ring-opened structures. When exciting to the $\tilde{A}$ state, it is thus expected that after 200~fs approximately 30 \% of the thiophene will be in a ring-open form. If excited to the $\tilde{B}$ state, however, approximately 60 \% will have ring-opened after 200~fs.

\section{Discussion and Conclusions}

In conclusion, by applying a high-order vibronic coupling model and the MCTDH method to the excited state dynamics of thiophene we have been able to assign the long unassigned vibronic features in the ultraviolet absorption spectrum of thiophene. The agreement between the calculated and measured spectrum is excellent, requiring only a small shift of 0.05 eV to align the first peak.
Both the spectral features and width are in near quantitative agreement.
Given that all parameters used in the model, including vertical excitation energies, were taken directly from the \emph{ab initio} calculations supports the quality of the CASPT2 treatment taken from the earlier work of Schnappinger et al. \cite{Schnappinger2017} 

The spectra due to the diabatic $\tilde{A}$ and $\tilde{B}$ states overlap
strongly, with the start of the $\tilde{B}$ band only 0.1 eV higher in energy than the $\tilde{A}$ even though the vertical excitation is 0.4 eV higher.
The spectral features arise due to the vibrational and vibronic coupling of a number of modes in the S$_0$ - S$_4$ manifold. The four vibrations $\nu_1, \nu_2, \nu_3$ and $\nu_7$ are responsible for the narrowly spaced triplet of peaks on the rising edge of the spectrum, while the combination of $\nu_{11}$ and $\nu_{13}$ provides the progression repeating this feature to higher energies. This progression has previously been assigned to the $\nu_6$ vibration. It is also found that the initial small peaks below the first sharp peak at 5.15eV are also due to the coupled vibrations and not due to hot bands as previously thought. This brings the origin of the band down in energy to just above 5.1 eV.

The quality of the spectrum means that the vibronic coupling model is a good description of the potential surfaces and couplings around the Franck-Condon point. It can thus also be used to gain
insight in the photo-excited dynamics of thiophene. 
While the $\tilde{C}$ and $\tilde{D}$ states are dark, they both play a role to in the dynamics, and indeed the features of the spectrum require the inclusion of these states in the model. The importance of these states is also seen in a recent study of the 1+1 REMPI spectra. \cite{bro24:xxx} The measured photoelectron angular distributions (PADs)
indicate that after photo-excitation to this band electrons are emitted from orbitals with $\sigma$ character which indicates crossing occurs to these two states from the initially populated $\tilde{A}$ and $\tilde{B}$.

The bright excited-states have a very different population dynamics.
If excited into the lower, $\tilde{A}$, state the initial population is 
found to relax fairly slowly, decreasing by approximately 40 \% over 200 fs, and driven initially by the $\nu_{16}$ vibration. This final population is spread fairly evenly over the ground state and  $\tilde{C}$ and $\tilde{D}$. In contrast, excitation to the $\tilde{B}$ state results in fast relaxation to $\tilde{A}$ followed by further crossing to $\tilde{X}, \tilde{C}$ and $\tilde{D}$. The initially populated state is depleted after 100 fs. This very different behaviour explains the different dynamics seen by Weinkauf {\em et al} \cite{Weinkauf2008} and Schalk {\em et al} \cite{Schalk2018} who excited into the $\tilde{A}$ and $\tilde{B}$ bands, respectively.

One of the largest controversies in the photochemistry of thiophene is whether, following photoexcitation, it ring opens or instead puckers out of plane. There is no clear experimental evidence for either, although
the resonance Raman study by Wu \emph{et al} expected ring opening to occur due to the importance of the observed anti-symmetric C-S-C stretch.\cite{res_raman} From the analysis of the state population dynamics and structures related to the minima on the potential surfaces the simulations clearly show that both are likely to happen. The ring puckering is due to the vibronic activation of the out-of-plane $\nu_1$ and $\nu_2$ 
of the $\tilde{A}$ state will take these geometries. In contrast, in-plane ring opening happens due to vibronic activation of the anti-symmetric C-S-C vibration $\nu_5$ leading to crossing from $\tilde{A}$ to $\tilde{B}$ to the ring open states $\tilde{C}$ and $\tilde{D}$. Exact time scales or yields for each will be difficult to extract due to the strong coupling of all the modes and states involved, but the ring-opening is expected to occur on a 100~fs timescale and increase in importance with excitation energy as crossing to the higher states becomes easier.

The second controversy in the photochemistry of thiophene is whether triplet states are involved or not. The question arises due to the presence of the sulphur atom suggesting a relatively large spin-orbit coupling and the importance of triplet states in the photochemistry of bithiophenes (and larger conjugated systems). Though thiophene has been found to demonstrate a very weak phosphorescence it is unclear how important triplets are to the dynamics following photoexcitation\cite{Becker_phos}. As we did not include spin-orbit coupling, the work presented here is not able to answer this question. Certainly our calculations show that triplet electronic states are not required to explain thiophene's absorption spectrum and due to the strong vibronic coupling are unlikely to play a significant role in the excited state dynamics of the first few 100~fs. Schnappinger \emph{et al} used surface hopping calculations to look at the interplay between intersystem crossing and internal conversion and came to conclusion that the triplet states were important in controlling how long it took thiophene to move from the ring open form back into the closed form.\cite{Schnappinger2017} However, like the majority of previous theoretical work on thiophene the $\tilde{D}$ state was not included in the work. As we have demonstrated, this state plays a significant role in the early dynamics after photo-excitation. Therefore, to confirm the involvement of triplet states it will be necessary to repeat these calculations including this $^1A_2$ state and the triplet states.

In conclusion, the work presented here provides an assignment and coherent explanation for the structures seen in the ``simple'' UV absorption spectrum of thiophene. 
The work thus provides a platform for further understanding of this molecule by elucidating the main ingredients required to describe the excited-state dynamics of this molecule.
The absorption of a photon by thiophene in this energy region is seen to lead to a complicated dynamics involving four strongly coupled electronic states and eight vibronically and vibrationally coupled modes. Fast population transfer (\textless{}100 fs) between the states is seen along with both ring puckering and ring opening. Future work will be needed to look at question such as the involvement of triplet states and the long-time dynamics, but that is beyond the capabilities of the vibronic coupling model used here.

\section*{Data Availability Statement}

The data that supports the findings of this study are available within the article and its appendices. This includes the coordinates of the optimised ground state structure, pictures of the vibrational normal modes, cuts through the vibronic coupling model potentials and the wavefunction tree and basis sets used in the ML-MCTDH calculations. Details are also given of available datasets containing the files from the simulations presented \cite{data}.

\section*{Acknowledgments}
The work was funded in part by the EPSRC programme grant EP/V026690/1.


\bibliography{thio} 

\clearpage
\newpage
\appendix

\section{Optimised Geometry}

\vspace*{5cm}
\begin{table}[h]
    \centering
    \caption{Optimised structure of Thiophene using CAS with a (10,9) active space and a 6-31G* basis set}
    \label{tab:Sup_thio_struc}
    \begin{tabular}{c|ccc}\hline
        \multirow{2}{*}{Atom} & \multicolumn{3}{l}{Coordinates /\r{A}} \\
                              & x             & y            & z            \\ \hline
                      C       &      0.00000         &    -1.25219  &  0.02820      \\
                      C       &      0.00000         &    -0.71665  &  1.28096      \\
                      C       &      0.00000         &     0.71665  &  1.28096      \\
                      C       &      0.00000         &     1.25219  &  0.02820      \\
                      S       &      0.00000         &     0.00000  & -1.22248      \\
                      H       &      0.00000         &    -2.28656  & -0.24758     \\
                      H       &      0.00000         &    -1.31280  &  2.17338     \\
                      H       &      0.00000         &     1.31280  &  2.17338     \\
                      H       &      0.00000         &     2.28656  &  -0.24758     \\\hline

    \end{tabular}
\end{table}

\clearpage
\newpage
\section{Vibronic Coupling Model}
\subsection{Vibrational Analysis}

\begin{figure}[h!]
	\unitlength1cm
    \begin{picture}(10,18.5)
    \put(-2,0.5){\includegraphics[scale=0.9]{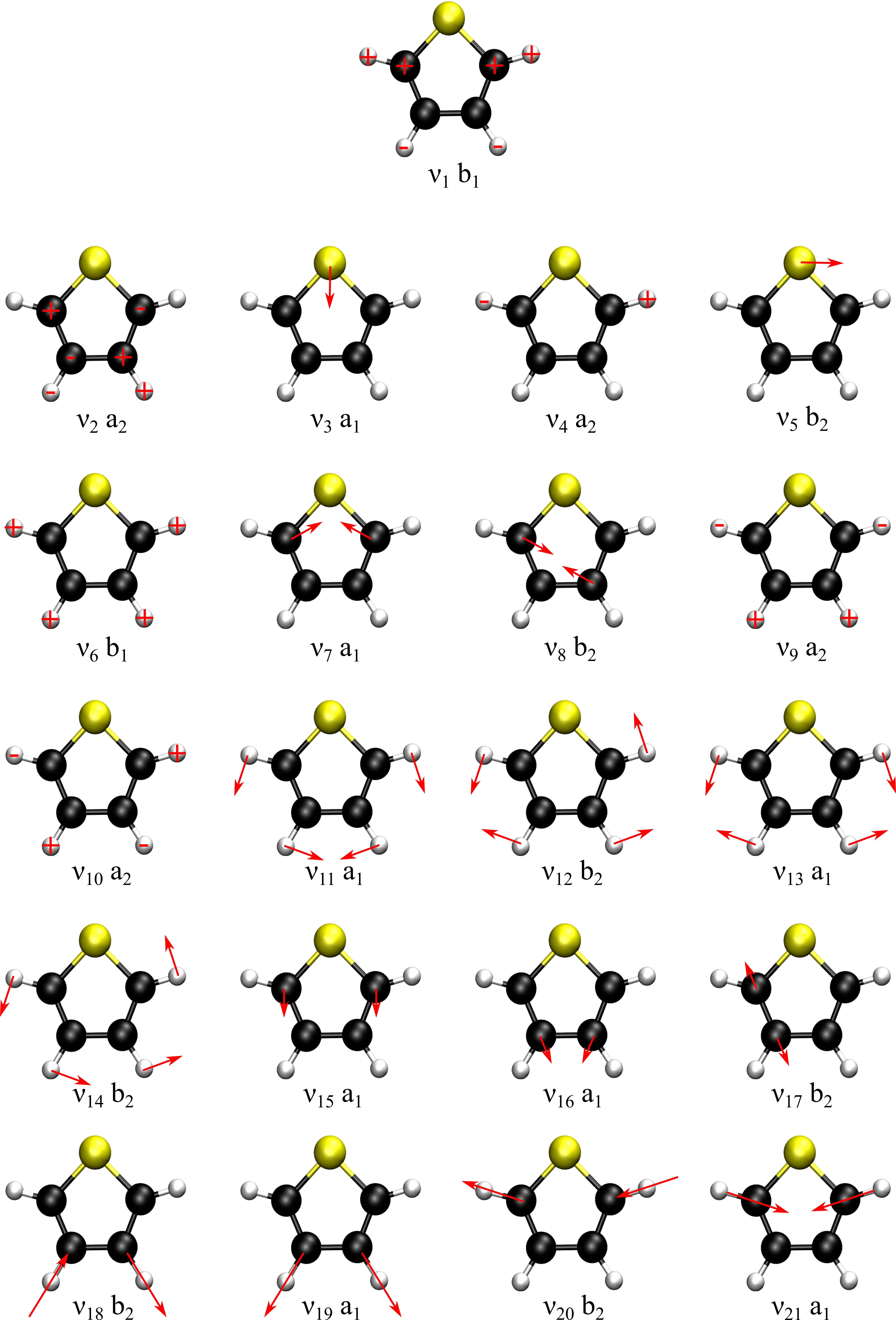}}
    \end{picture}
    \caption{Representation of the normal modes of thiophene calculated using SA4-CAS(9,10)/6-31G*.}
    \label{fgr:supMOs}
\end{figure}

\clearpage
\newpage
\subsection{Parameters}

\begin{table}[h!]
\caption{Significant $\kappa_\alpha^{(i)}$ parameters in the Vibronic Coupling Hamiltonian of thiophene for mode $\alpha$ in state $i$. Only values with a coupling strength greater than 0.2 listed. All values are in eV. }
\centering
    \begin{tabular}{lrrrrr} \hline
            & \multicolumn{4}{c}{State $(i)$} \\
                    &    2    &   3    &    4    &    5   \\\hline
$\kappa^{(i)}_{3}$   & -0.039 &  0.078 &   0.309 &   0.232 \\
$\kappa^{(i)}_{7}$   &  0.038 & -0.057 &   --    &   0.123 \\
$\kappa^{(i)}_{11}$  &  0.102 & -0.028 & -0.031  &  0.094 \\
$\kappa^{(i)}_{13}$  &  0.087 &  0.156 &  0.153  &  0.048 \\
$\kappa^{(i)}_{15}$  &  0.099 &  0.136 &  0.042  &   -- \\
$\kappa^{(i)}_{16}$  & -0.174 & -0.347 & -0.139  &  0.223 \\
    \hline
    \end{tabular}
\label{tab:kappa}
\end{table}

\begin{table}[h!]
\caption{Significant $\lambda_\alpha^{(ij)}$ parameters in the Vibronic Coupling Hamiltonian of thiophene for mode $\alpha$ coupling states $i$ and $j$. Only values with a coupling strength greater than 0.2 listed. All values are in eV. }
\centering
    \begin{tabular}{rcrrrr|rcrrrr} \hline
	    &  & \multicolumn{4}{c|}{State $(j)$} & &  & \multicolumn{4}{c}{State $(j)$} \\
  &  State $(i)$  &    2    &   3    &    4    &    5   &
  &  State $(i)$  &    2    &   3    &    4    &    5   \\\hline
 $\lambda^{(i,j)}_{1}$   &  1 & & &  -0.099 & &   $\lambda^{(i,j)}_{8}$   &  1 & &  0.174 & &   \\ 
                         &  2 & & & -0.024 &  &                           &  4 & & & &  -0.040  \\ 
                         &  3 & & & &  -0.034 &   $\lambda^{(i,j)}_{9}$   &  1 & & & -0.05 &    \\
 $\lambda^{(i,j)}_{2}$   &  1 & & & & 0.143   &   $\lambda^{(i,j)}_{10}$  &  1 & & & & 0.042    \\
                         &  2 & & & & 0.100   &                           &  2 & & & & 0.041    \\
 $\lambda^{(i,j)}_{3}$   &  1 & -0.120 & & &  &                           &  3 & & &  0.030 &   \\ 
 $\lambda^{(i,j)}_{4}$   &  1 & & & & 0.049   &   $\lambda^{(i,j)}_{11}$  &  1 & -0.229 & & &   \\
                         &  3 & & &  -0.049 & &   $\lambda^{(i,j)}_{12}$  &  4 & & & &  -0.053  \\ 
 $\lambda^{(i,j)}_{5}$   &  1 & & -0.170 & &  &   $\lambda^{(i,j)}_{13}$  &  1 &  0.114 & & &   \\ 
                         &  2 & & -0.102 & &  &   $\lambda^{(i,j)}_{14}$  &  1 & & &  0.027 &   \\ 
                         &  4 & & & &  0.093  &                           &  2 & & 0.038 & &    \\
 $\lambda^{(i,j)}_{6}$   &  1 & & &  -0.079 & &   $\lambda^{(i,j)}_{15}$  &  1 &  0.041 & & &   \\ 
                         &  2 & & &  0.032 &  &   $\lambda^{(i,j)}_{16}$  &  1 & -0.334 & & &   \\ 
                         &  3 & & & &  -0.019 &   $\lambda^{(i,j)}_{17}$  &  4 & & & &  0.108   \\
 $\lambda^{(i,j)}_{7}$   &  1 & -0.105 & & &  &  & & & & \\ 
    \hline
    \end{tabular}
\label{tab:lambda}
\end{table}

\begin{table}[h!]
\caption{High-Order expansion terms added to the LVC model model Hamiltonian to provide a fit of the adiabatic surfaces. The first block of terms are for the diabatic potentials along a single mode, $x$. The second block provide terms in a diabatic potential that couple modes $x$ and $y$. The third block couple states $i$ and $j$ as well as modes $x$ and $y$. Mode numbering is as in Table II in the main paper. }
\label{tab:VCterms}
\centering
    \begin{tabular}{lll|lllll} \hline
   Terms & \multicolumn{1}{c}{Mode} & \multicolumn{1}{c|}{State} &
   Terms & \multicolumn{2}{c}{Mode} & \multicolumn{2}{c}{States} \\
	    &    x   &    i   &   &    x   &   y   &  i  & j \\\hline
    $a_4 x^4$ & 1 & 1-3, 5 &                     $(b_2 x + b_3 x^2 +b_4 x^3 + b_5 x^4) y$ & 3 & 5 & 1 &                \\
    $a_4 x^4 +a_6 x^6$ & 1 & 4 &                 $(b_2 x + b_3 x^2 +b_4 x^3 + b_5 x^4 +b_7 x^6) y$ & 3 & 5 & 2-5 &     \\
    $a_4 x^4$ & 2 & 1, 3, 4 &                    $b_4 x^2 y^2$ & 1 & 2 & 1-5 &                                         \\
    $a_4 x^4 +a_6 x^6$ & 2 & 2, 5 &              $x^2 (b_3 y + b_4 y^2 + b_5 y^3 + b_6 y^4 )$ &  1 & 3 & 2-5 &         \\
    $a_3 x^3 + a_4 x^4 +a_6 x^6$ & 3 & 1-5 &     $x^2 (b_3 y + b_4 y^2 + b_5 y^3 + b_6 y^4 )$ &  2 & 3 & 4,5 &         \\
    $a_4 x^4$ & 4 &  1-5 &                        $b_6 x^2 y^4$ &  2 & 7 & 2,4 &                                        \\
    $a_4 x^4$ & 5 &  2, 4, 5 &                    $x^2 (b_5 y^3 + b_6 y^4)$ &  2 & 7 & 5 &                              \\
    $a_4 x^4$ & 6 &  1-3 &                        $ b_6 x^2 y^4$ & 2 & 3 & 2 &                                          \\
    $a_3 x^3 + a_4 x^4 +a_6 x^6$ & 7  & 1-5 &     $ b_6 x^2 y^4$ & 1 & 3 & 1 &                                          \\
    $a_4 x^4$ & 8 &  3-5 &                        $x (b_2 y + b_3 y^2 +b_4 y^3 + b_5 y^4)$ & 5 & 7 & 1-5 &              \\
    $a_4 x^4$ & 9 &  2, 3 &                       $b_2 x y + b_{31} x^2 y + b_{32} x y^2 + b_4 x^2 y^2$ & 3 & 7 & 1-5 & \\
    $a_4 x^4$ & 10 &  2-5 &                       $x^2 (b_3 y + b_4 y^2 + b_5 y^3 + b_6 y^4 )$ &  1 & 7 & 2, 3 &        \\
    $a_3 x^3 + a_4 x^4$ & 11  & 1-5 &             $ b_3 x^2 y$ & 2 & 7 & 1 & 2                                          \\
    $a_4 x^4$ & 14 &  4 &                         $ b_3 x^2 y$ & 1 & 3 & 1 & 2                                          \\
    $a_3 x^3 + a_4 x^4$ & 15  & 1-5 &             $b_2 xy$ & 5 & 7 & 2 & 3                                              \\
    $a_3 x^3 + a_4 x^4$ & 16  & 1-3 &             $b_2 xy$ & 5 & 7 & 4 & 5                                              \\
    $a_3 x^3 + a_4 x^4$ & 17  & 4, 5 &            $b_2 xy$ & 3 & 5 & 4 & 5                                              \\
    $a_3 x^3 + a_4 x^4$ & 19  & 1-5 &             $b_2 xy$ & 2 & 3 & 1 & 5                                              \\
    $a_3 x^3 + a_4 x^4$ & 21  & 1-5 &       & & & &   \\ 
    \hline
    \end{tabular}
\end{table}

\clearpage
\newpage
\subsection{Cuts Through the Potential Surfaces}

\begin{figure}[h!]
	\unitlength1cm
    \begin{picture}(10,15)
    \put(-3,0.5){\includegraphics[scale=1.16]{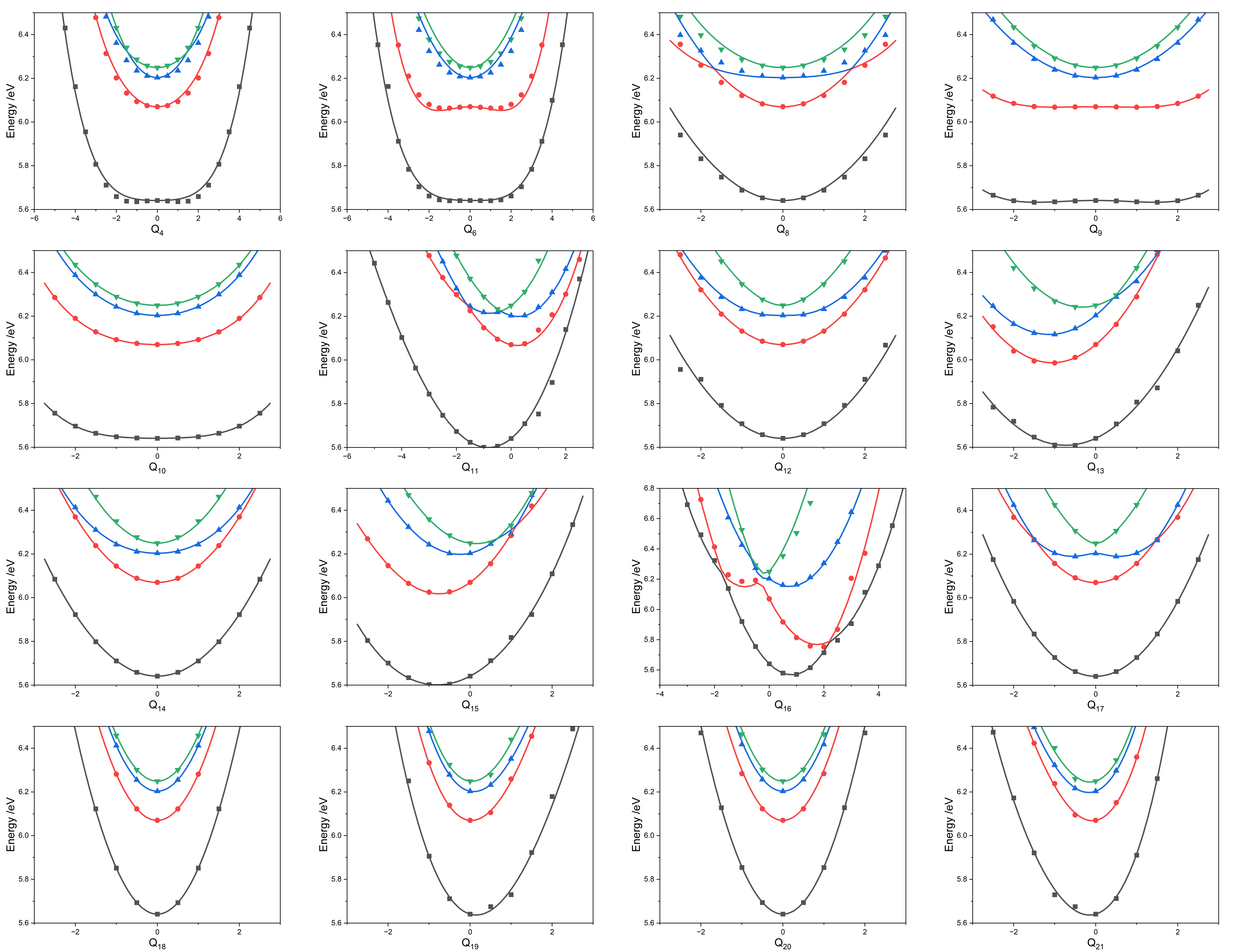}}
    \end{picture}
    \caption{Cuts through the Thiophene potential energy surface along the normal modes calculated using CAS-PT2 with a (9,10) active space and a 6-31G* basis set. The x axes are the mass-and frequency-scaled normal mode coordinates. Symbols are calculated energies while solid lines are the fits to the surfaces from VCHAM. Only states S$_1$ to S$_4$ are shown }
    \label{fig:supCurves}
\end{figure}

\clearpage
\newpage
\subsection{Minima on the S$_1$ Adiabatic Surface}

\begin{table}[h!]
    \caption{Mass-frequency scaled normal mode coordinates of the two minima found on the S$_1$ adiabatic potential surface of the thiophene vibronic coupling model Hamiltonian.}
    \label{tab:Qmin_S1}
    \centering
\begin{tabular}{rc|rr}\hline
    &  Symmetry&   Ring Pucker     &  Ring Open  \\ \hline
  1 &  B$_1$   &    5.856  &  -0.037  \\ 
  2 &  A$_2$   &    0.900  &  -1.432  \\ 
  3 &  A$_1$   &    0.588  &  -7.392  \\ 
  4 &  A$_2$   &   -0.364  &   0.103  \\ 
  5 &  B$_2$   &   -0.066  &   9.596  \\ 
  6 &  B$_1$   &   -0.683  &   0.198  \\ 
  7 &  A$_1$   &    0.293  & -11.180  \\ 
  8 &  B$_2$   &   -0.003  &  -0.584  \\ 
  9 &  B$_1$   &    1.400  &   0.051  \\ 
 10 &  A$_2$   &    0.356  &  -0.086  \\ 
 11 &  A$_1$   &   -1.264  &  -0.272  \\ 
 12 &  B$_2$   &   -0.014  &   0.038  \\ 
 13 &  A$_1$   &   -0.640  &  -0.570  \\ 
 14 &  B$_2$   &    0.015  &  -0.253  \\ 
 15 &  A$_1$   &   -0.755  &  -0.165  \\ 
 16 &  A$_1$   &    0.775  &   0.142  \\ 
 17 &  B$_2$   &    0.001  &  -0.009  \\ 
 18 &  B$_2$   &    0.001  &   0.005  \\ 
 19 &  A$_1$   &    0.143  &   0.029  \\ 
 20 &  B$_2$   &   -0.000  &   0.016  \\ 
 21 &  A$_1$   &   -0.158  &  -0.085  \\ \hline
\multicolumn{2}{c}{Energy}    &  5.231   &  2.772    \\\hline
 \end{tabular}
 \end{table}

 \begin{table}[h!]
    \caption{Cartesian coordinates (in {\AA}) of the two minima found on the S$_1$ adiabatic potential surface of the thiophene vibronic coupling model Hamiltonian.}
    \label{tab:Xmin_S1}
    \centering
\begin{tabular}{lrrrclrrr}\cline{1-4}\cline{6-9}
  \multicolumn{4}{c}{Ring Pucker} & \hspace*{2cm} & \multicolumn{4}{c}{Ring Open} \\\cline{1-4}\cline{6-9}
   C  &   0.256 &  -1.263 &  -0.017   &  &  C  &   0.034 & -1.743 &  0.460 \\
   C  &  -0.096 &  -0.731 &   1.303   &  &  C  &  -0.049 & -0.862 &  1.581 \\
   C  &  -0.157 &   0.733 &   1.306   &  &  C  &   0.045 &  0.660 &  1.211 \\
   C  &   0.291 &   1.265 &  -0.014   &  &  C  &  -0.040 &  1.415 &  0.060 \\
   S  &  -0.112 &  -0.001 &  -1.202   &  &  S  &   0.001 &  0.201 & -1.529 \\
   H  &   0.117 &  -2.294 &  -0.350   &  &  H  &   0.020 & -2.832 &  0.311 \\
   H  &  -0.032 &  -1.360 &   2.196   &  &  H  &  -0.104 & -1.010 &  2.792 \\
   H  &  -0.242 &   1.361 &   2.198   &  &  H  &   0.154 &  1.205 &  2.148 \\
   H  &   0.231 &   2.295 &  -0.348   &  &  H  &   0.007 &  2.544 &  0.073 \\\cline{1-4}\cline{6-9}
 \end{tabular}
 \end{table}

 \clearpage
\newpage
\subsection{Minima on the S$_2$ Adiabatic Surface}

\begin{table}[h!]
    \caption{Mass-frequency scaled normal mode coordinates of the two minima found on the S$_2$ adiabatic potential surface of the thiophene vibronic coupling model Hamiltonian.}
    \label{tab:Qmin_S2}
    \centering
\begin{tabular}{rc|rrrr}\hline
    &  Symmetry&  Ring Pucker 1 &  Ring Pucker 2 &  Ring Pucker 3 &  Ring Open    \\ \hline
  1 &  B$_1$   &    7.465  &   4.336 &   4.669 &  -0.000 \\ 
  2 &  A$_2$   &   -0.000  &  -1.319 &   0.071 &  -0.000 \\  
  3 &  A$_1$   &   -1.189  &  -1.907 &  -1.856 &  -7.061 \\  
  4 &  A$_2$   &    0.000  &  -0.039 &   0.039 &   0.000 \\  
  5 &  B$_2$   &   -0.001  &   2.844 &  -0.159 &   9.317 \\  
  6 &  B$_1$   &   -0.052  &   0.145 &  -0.078 &   0.000 \\  
  7 &  A$_1$   &    2.134  &  -0.464 &  -1.191 &  -9.648 \\  
  8 &  B$_2$   &    0.001  &  -0.232 &   0.058 &  -0.340 \\  
  9 &  B$_1$   &    0.105  &  -0.003 &   0.046 &   0.000 \\  
 10 &  A$_2$   &   -0.000  &  -0.242 &   0.009 &   0.000 \\  
 11 &  A$_1$   &   -1.293  &  -0.606 &  -0.740 &  -0.492 \\  
 12 &  B$_2$   &    0.001  &  -0.025 &   0.043 &   0.420 \\  
 13 &  A$_1$   &   -0.625  &  -0.783 &  -0.687 &  -0.555 \\  
 14 &  B$_2$   &   -0.002  &  -0.200 &   0.010 &  -0.085 \\  
 15 &  A$_1$   &   -0.726  &  -0.489 &  -0.119 &   0.085 \\  
 16 &  A$_1$   &    0.766  &   0.684 &  -0.401 &  -0.500 \\  
 17 &  B$_2$   &    0.001  &   0.185 &  -0.061 &  -0.611 \\  
 18 &  B$_2$   &    0.000  &   0.019 &  -0.002 &  -0.012 \\  
 19 &  A$_1$   &    0.126  &   0.108 &   0.044 &   0.057 \\  
 20 &  B$_2$   &    0.000  &   0.006 &  -0.000 &   0.010 \\  
 21 &  A$_1$   &   -0.145  &  -0.176 &  -0.112 &  -0.135 \\  \hline
\multicolumn{2}{c}{Energy}   &   5.342  &   5.498 &   5.677 &   3.953     \\\hline
 \end{tabular}
 \end{table}

 \begin{table}[h!]
    \caption{Cartesian coordinates (in {\AA}) of the two minima found on the S$_2$ adiabatic potential surface of the thiophene vibronic coupling model Hamiltonian.}
    \label{tab:Xmin_S2}
    \centering
\begin{tabular}{lrrrclrrr}\cline{1-4}\cline{6-9}
  \multicolumn{4}{c}{Ring Pucker 1} & \hspace*{2cm} & \multicolumn{4}{c}{Ring Pucker 2} \\\cline{1-4}\cline{6-9}
   C  &    0.323 & -1.180 & -0.004 &  &  C  &   0.220 & -1.312 &  0.100 \\
   C  &   -0.148 & -0.726 &  1.353 &  &  C  &  -0.125 & -0.745 &  1.403 \\
   C  &   -0.148 &  0.726 &  1.353 &  &  C  &  -0.048 &  0.693 &  1.288 \\
   C  &    0.323 &  1.180 & -0.004 &  &  C  &   0.150 &  1.208 & -0.014 \\
   S  &   -0.143 & -0.000 & -1.247 &  &  S  &  -0.083 &  0.060 & -1.287 \\
   H  &    0.450 & -2.208 & -0.341 &  &  H  &   0.281 & -2.349 & -0.218 \\
   H  &   -0.243 & -1.424 &  2.196 &  &  H  &  -0.270 & -1.304 &  2.342 \\
   H  &   -0.243 &  1.424 &  2.197 &  &  H  &   0.010 &  1.339 &  2.163 \\
   H  &    0.450 &  2.208 & -0.341 &  &  H  &   0.289 &  2.260 & -0.276 \\\cline{1-4}\cline{6-9}
    & & & &  &  &  &  &  \\
    & & & &  &  &  &  &  \\\cline{1-4}\cline{6-9}
  \multicolumn{4}{c}{Ring Pucker 3} & \hspace*{2cm} & \multicolumn{4}{c}{Ring Open} \\\cline{1-4}\cline{6-9}
   C  &   0.199 & -1.274 &  0.065 &  &  C  &   0.000 & -1.671 &  0.454 \\
   C  &  -0.090 & -0.744 &  1.321 &  &  C  &   0.000 & -0.881 &  1.559 \\
   C  &  -0.093 &  0.744 &  1.328 &  &  C  &   0.000 &  0.660 &  1.217 \\
   C  &   0.204 &  1.283 &  0.069 &  &  C  &   0.000 &  1.377 &  0.044 \\
   S  &  -0.090 & -0.003 & -1.290 &  &  S  &   0.000 &  0.194 & -1.509 \\
   H  &   0.284 & -2.307 & -0.246 &  &  H  &   0.000 & -2.748 &  0.268 \\
   H  &  -0.149 & -1.293 &  2.255 &  &  H  &   0.000 & -1.040 &  2.762 \\
   H  &  -0.166 &  1.289 &  2.266 &  &  H  &   0.000 &  1.245 &  2.134 \\
   H  &   0.273 &  2.316 & -0.246 &  &  H  &   0.000 &  2.487 & -0.019 \\\cline{1-4}\cline{6-9}
 \end{tabular}
 \end{table}

\clearpage
\newpage
\section{ML-MCTDH Basis Set}

\begin{figure}[h!]
    \unitlength1cm
    \begin{picture}(5,9.5)
    \put(-6,0){\includegraphics[scale=0.6]{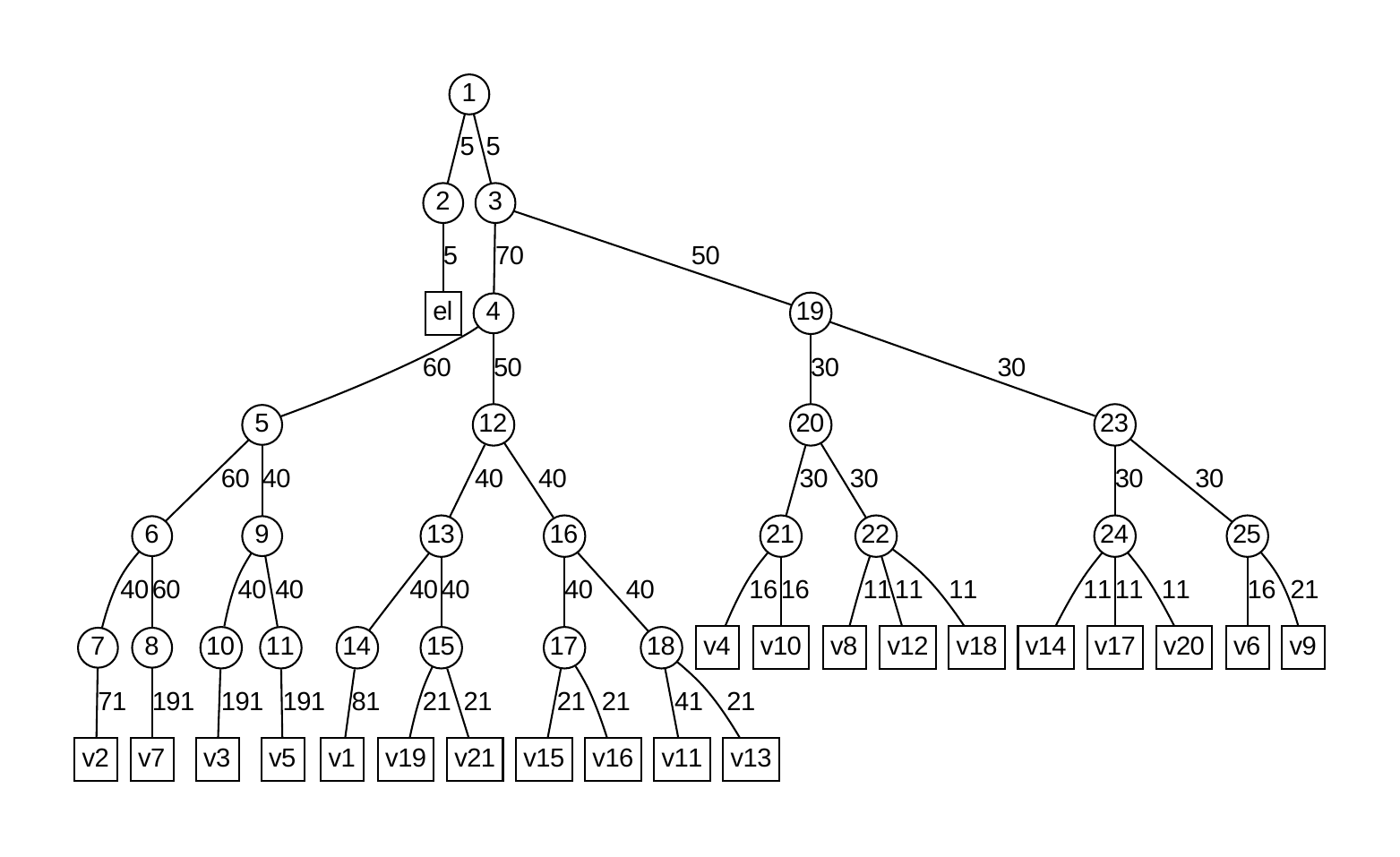}}
    \end{picture}
    \caption{The ML-MCTDH tree defining the structures of the layers and the number
        of basis functions used. The circles are the nodes represented by SPFs in
        the layers. The squares for the lowest layer represent the time-independent primitive
        basis. This was a harmonic oscillator DVR. The numbers on the lines connecting the
        nodes are the number of functions used.}
    \label{fig:sup_mltree}
\end{figure}

\clearpage
\newpage
\section{Comparison of Spectra}

\begin{figure}[h!]
    \unitlength1cm
    \begin{picture}(8,6.5)
    \put(-3,0){\includegraphics[scale=1.1]{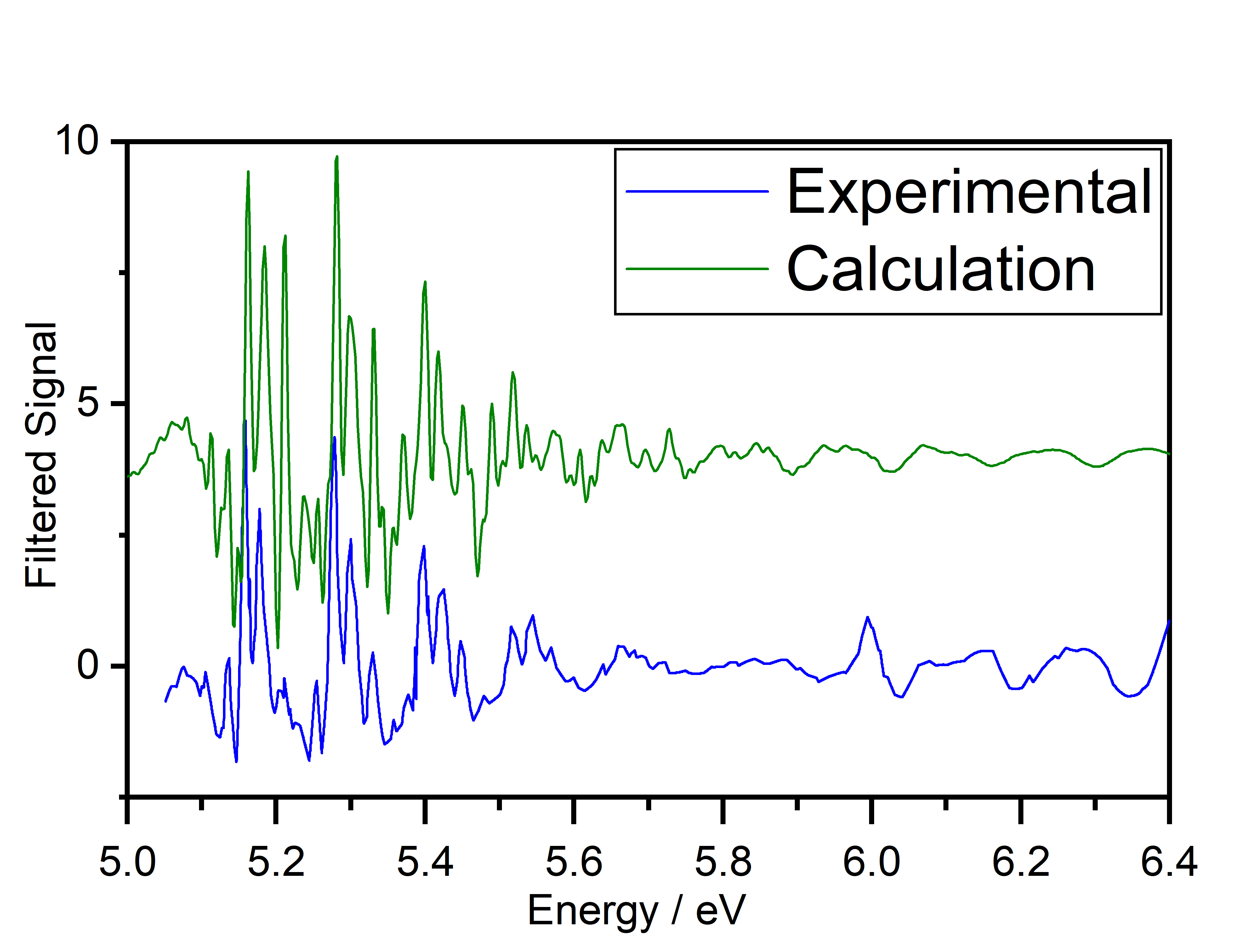}}
    \end{picture}
    \caption{Comparison of experimental and calculated photoabsorption spectra following application of a 7~Hz High pass Fast Fourier Transform filter. This filter reveals the sharp features obscured by the underlying broad spectrum. Experimental data taken from Holland \emph{et al}.\cite{Holland2014} The calculation used the full Hamiltonian. The plots have been offset from each other for clarity}
    \label{fig:sup_fft}
\end{figure}

\section{Data Available for the Quantics Calculations}

The input and operators files for the Quantics MCTDH calculations, along with the output files, are available on the UCL Research Data Repository at DOI: 10.5522/04/26076520. The files, including the electronic structure output, used in the fitting of the potential surfaces by the VCHAM program are also available at that site.

\end{document}